\documentclass[useAMS,usenatbib]{mn2e}
\usepackage{graphicx}

\title[A Scaling Index Analysis of the WMAP three year data]
     {A Scaling Index Analysis of the WMAP three year data: 
       Signatures of non-Gaussianities and Asymmetries in the CMB}

\author[C. R\"ath, P.Schuecker and A. J. Banday]
       {C. ~R\"ath,$^1$\thanks{E-mail: cwr@mpe.mpg.de}   P. ~Schuecker$^1$ and A. J. ~Banday$^2$  \\
        $^1$ Max-Planck Institut f\"ur extraterrestrische Physik,
        Giessenbachstr. 1, 85748 Garching, Germany\\
         $^2$ Max-Planck Institut f\"ur Astrophysik, Karl- Schwarzschildstr.1, 
         85741 Garching, Germany}

\begin{document}

\date{Draft version, 6 February 2007 }
\pagerange{\pageref{firstpage}--\pageref{lastpage}} \pubyear{2007}

\maketitle

\label{firstpage}

\begin{abstract}

Local scaling properties of the co-added foreground-cleaned 
three-year { \it Wilkinson Microwave Anisotropy Probe (WMAP)} 
data are estimated using weighted scaling indices $\alpha$.
The scaling index method (SIM) is - for the first time - adapted and applied    
to the case of spherical symmetric spatial data.
The results are compared with $1000$ Monte Carlo simulations
based on Gaussian fluctuations with a best fit $\Lambda$CDM power spectrum and
WMAP-like beam and noise properties.
Statistical quantities based on the scaling indices, namely
the moments of the distribution and  probability-based measures are determined.
We find for most of the test statistics significant deviations from the Gaussian hypothesis.  
Using a very conservative $\chi^2$ statistics, which averages over all 
scales, we detect non-Gaussianity with a probability 
of $97.3$ \% regarding the Kp0-masked full sky,  $98.9$  \% for the Kp0-masked northern and $91.6$  \% 
for the Kp0-masked southern hemisphere. When analysing different length scales separately, 
the detection rates increase to  $99.7$ \% ($>99.9$ \% north, $97.4$ \% south) 
for the mean, $98.4$ \% ($99.9$ \% north, $71.6$ \% south) for the standard deviation 
and $97.9$ \% ($99.4$ \% north, $80.3$ \% south) for a  $\chi^2$-combination of  
mean and standard deviation.
We find pronounced asymmetries, which can be interpreted as a global lack of structure 
in the northern hemisphere, which is consistent with previous findings.
Furthermore, we detect a localized anomaly in the southern hemisphere, which gives rise 
to highly significant signature for non-Gaussianity in the spectrum of scaling indices $P(\alpha)$. 
We identify this signature as the cold spot, which was also already detected 
in the first year WMAP  data. 
Our results provide further evidence for both the presence of non-Gaussianities  
and asymmetries in the WMAP three-year data.
More detailed band- and year-wise analyses are needed to elucidate the origin of the 
detected anomalies. In either case the scaling indices provide powerful 
nonlinear statistics to analyse CMB maps.

\end{abstract}


\section{Introduction}

The study of the Gaussianity of the Cosmic Microwave Background (CMB) is regarded 
to be the best way for understanding the true nature of the primordial density fluctuations: 
by measuring suitable statistics of the temperature fluctuations of the CMB 
and comparing the results with theoretical predictions, 
cosmological models for the primordial density fluctuations can be rejected 
or accepted at a certain confidence level. 
Standard inflationary models \citep{guth81a,linde82a,albrecht82a} 
predict that the temperature fluctuations of the CMB 
correspond to a (nearly) Gaussian, homogeneous and isotropic random field.
In fact, Gaussianity and statistical isotropy  are among the fundamental pillars 
of the $\Lambda$CDM concordance cosmological model.
On the other hand, many alternative scenarios have been studied, which give rise
to non-Gaussianity, e.g. non-standard 
inflation \citep{linde97a,peebles97a,bernardeau02a,acquaviva03a},
or topological defects models \citep{bouchet88a, turok90a, turok96a}
to mention a few.
Probing non-Gaussianity therefore represent one of the key tests to detect 
deviations from the minimal scenario.
To detect them it is essential to have high sensitivity 
and large-coverage CMB data.\\  
The {\it Wilkinson Microwave Anisotropy Probe (WMAP)} satellite has
produced high resolution all-sky observations
of the CMB with unprecendented accuracy, which have
confirmed many  predictions of the   $\Lambda$CDM 
concordance cosmological model \citep{spergel03a, bennett03a, spergel06a}.\\ 
First analyses based on the WMAP first year data  and 
global statistical measures  
did not show significant deviations from 
the Gaussianity hypothesis \citep{komatsu03b}. 
A number of  subsequent studies of the  first-year WMAP-data  yielded, however,
evidence for non-Gaussian features  as well as asymmetries in the CMB 
as measured with the WMAP data.\\
These results were obtained using a number of qualitatively different 
statistical tests ranging from  (local) spectral \citep{deoliveira04a, copi04a,eriksen04b,hansen04a,land05a} 
and $N$-point correlation  function analysis \citep{eriksen04b,eriksen05a},
 to Fourier phases \citep{chiang03a, coles04a, naselsky05a}, 
 Minkowski - functionals \citep{park04a,eriksen04a} 
 and wavelets \citep{vielva04a,cruz05a,mcewen05a}.\\
In March 2006 the three-year WMAP results were released and
several groups revisited the anomalies found in the first data release.
Considering the fact that the first year WMAP data were already of very high quality
it was no surprise  that  
many of the anomalies found in the first year data were 
redetected \citep{bridges06a,copi06b,jaffe06a, land06a,martinez06a} in the 
three year data release albeit
none of these tests alone  showed outstanding significance.
Hence there is an ongoing discussion, whether the identified features are 
significant enough to prove asymmetry or non-Gaussianity of the CMB.
Given the tremendous consequences a claim of non-Gaussianity of the CMB on cosmology would have,  
some caution in making such statements is warranted.\\   
Non-Gaussianity cannot be  defined in a unique manner, 
it is simply the presence of any higher order spatial correlations.  
Therefore, there cannot be one unique test for all possible ways 
a random field can be non-Gaussian that is preferable to other tests, 
or that has superior sensitivity. 
In order to perform a thorough analysis, one rather has to apply several different, 
preferably complementary tests.
Following this reasoning, we present in this paper a novel approach to analyse 
observed CMB-data on a local scale  and to test for Gaussianity and (statistical) isotropy. 
Namely we estimate the local scaling properties of the maps of temperature fluctuations 
using weighted scaling indices (see e.g. \citet{raeth02a,raeth03a}).
These scaling indices are sensitive to the local morphological properties of the field of
the temperature fluctuations $\Delta T(\theta, \phi)$ at a given scale $r$.
It has been demonstrated that this statistic is highly sensitive to
non-Gaussian signatures in simulated CMB-maps. 
Here,  we propose 
a formalism that can be applied to observed WMAP-data with its spherical symmetry and
test for both non-Gaussianties and statistical asymmetries in the WMAP three-year data 
using Monte Carlo simulations based on Gaussian fluctuations with a best fit $\Lambda$CDM power spectrum and
WMAP-like beam and noise properties.\\
The paper is organised as follows.  
In Section 2 we describe the WMAP-data as well as the simulations and 
the preprocessing steps, which were applied to both the simulated and observed data.  
In Section 3 we review the formalism of
the weighted scaling indices and explain it with a simple example.  
As the weighted scaling indices and the wavelets share some properties 
(e.g. locality, scale dependence) we compare the two test statistics, 
show similarities and outline differences.
We further extend the scaling index formalism to the application  
to spherical WMAP-data.
In Section 4  the results of the WMAP three-year data analysis are presented. 
We summarize  our main results in Section 5 and present our conclusions in Section 6.   

\section{WMAP Data, Simulations and Preprocessing}

We use the noise-weighted 
sum $T$ of the  V1, V2, W1, W2, W3 and W4 foreground-cleaned maps,
\begin{equation}
T(\theta, \phi) = \frac{\sum_{j=5}^{10} T_j(\theta,\phi)/ \sigma^2_{0,j} } {\sum_{j=5}^{10} 1/ \sigma^2_{0,j}}
\end{equation}
where $j=5,6$ refer to the two V-band receivers and $j=7,\ldots,10$ to the four W-bands.
$\sigma^2_{0,j}$ denotes the  three-year noise per observation for the six frequency 
bands given by \citet{hinshaw06a}.  
In contrast to many of the  one-year Gaussianity studies (e.g.  \citet{komatsu03b, vielva04a}) 
the Q bands were no longer taken into account, because they exhibit foreground contaminations.
All these foreground cleaned maps were taken from the publicly accessible 
LAMBDA website\footnote{http://lambda.gsfc.nasa.gov}. \\
In order to study the Gaussianity of the coadded VW-WMAP maps, we generated and analysed a set of 
1000 Gaussian Monte Carlo simulations. 
For this  we took the set of the best fit $C_l$'s  (WMAP only) 
as published on the LAMBDA-site some weeks after
the three year data release
\footnote{http://lambda.gsfc.nasa.gov/product/map/current/params/lcdm\_wmap.cfm}. 
For the set of $C_l$'s, random $a_{lm}$'s for the $C_l$'s of CMB realizations were generated 
and convolved at each one of the WMAP receivers with the appropriate beam transfer functions
\footnote{http://lambda.gsfc.nasa.gov/product/map/dr2/xfer\_funcs\_get.cfm}. 
After the transformation from harmonic to real space, uncorrelated Gaussian 
noise realizations were added following the number of observations 
per pixel $N_j(\theta,\phi)$ and the noise dispersion per observation ($\sigma_{0j}$). 
We combined all the maps following equation (1). 
Finally, both the coadded WMAP-data and the simulations 
were degraded to $N_{side} = 256$, the Kp0 mask was applied, 
and the residual monopole and dipoles were fitted and subtracted. 
All these simulation and preprocessing steps were performed 
with the HEALPIX-software \citep{gorski05a}.
  
\section{Weighted Scaling Indices}

The weighted scaling index method (SIM) \citep{raeth02a,raeth03a}  
offers one possibility to
estimate the {\it local} scaling properties of a set of points,
which is generally repesented in a $d-$dimensional space.
The SIM has found applications in (astrophysical)
time series analysis of AGNs, where the observed light curves were
represented as point distributions by embedding
the time series in a higher dimensional artificial phase space using
delay coordinates \citep{gliozzi02a, gliozzi06a}.\\â
Scaling indices have also often successfully been used for structure analysis in 2D and 3D image data
\citep{jamitzky01a, monetti03a, mueller04a}. 
For this the image data are represented as point distributions by comprising the 
spatial and intensity information of each pixel. 
For two-dimensional images one thus obtains a set of 
three-dimensional vectors $\vec{p_{i}} = (x_i,y_i,I(x_i,y_i))$,$i=1,\ldots,N_{pixels}$.   
On the basis of the representation of images as point distributions the
weighted scaling indices are calculated as follows:

\subsection{General Formalism}
Consider a  set of $N$
points $P=\{\vec{p_i}\}, i=1,\ldots,N$.
For each point the local weighted cumulative point
distribution $\rho$ is calculated.
In general form this can be written as
\begin{equation}
  \rho(\vec{p_i},r) = \sum_{j=1}^{N} s_r (d(\vec{p_i},\vec{p_j})) \;,
\end{equation}
where $s_r(\bullet)$ denotes a kernel function depending on
the scale parameter $r$ and $d(\bullet)$ a distance measure.\\
The weighted scaling indices $\alpha(\vec{p_i},r)$ are obtained by calculating
the logarithmic derivative of $\rho(\vec{p_i},r)$ with respect to $r$,
\begin{equation}
  \alpha(\vec{p_i},r) = \frac{\partial \log \rho(\vec{p_i},r)}{\partial \log r}
                      = \frac{r}{\rho}\frac{\partial}{\partial r} \rho(\vec{p_i},r) \;.
\end{equation}
In principle, any differentiable kernel function and any distance measure can
be used for calculating $\alpha$.
In the following we use the euclidean norm as
distance measure and a set of gaussian shaping functions.
So the expression for $\rho$ simplifies to
\begin{equation}
  \rho(\vec{p_i},r) = \sum_{j=1}^{N} e^{-(\frac{d_{ij}}{r})^q} \;,
                       d_{ij} = \| \vec{p_i} - \vec{p_j} \| \;.
\end{equation}
The exponent $q$ controls the weighting of the points according to their
distance to the point for which $\alpha$ is calculated.
The higher $q$ is the more steplike becomes the weighting function resembling 
more and more the Heaviside-function, which is used for the 
calculation of the unweighted scaling indices.
Another interesting choice of $q$ is given by $q=2$. In this case the kernel function 
is the well-known Gaussian exponential function.\\ 
Throughout this study we calculate $\alpha$ for the case
$q=2$. 
Inserting the expression (4) in  the definition for the weighted scaling indices in (3) yields after some algebra
the following analytical expression for $\alpha$:
\begin{equation}
  \alpha(\vec{p_i},r) = \frac{\sum_{j=1}^{N} q (\frac{d_{ij}}{r})^q
                                             e^{-(\frac{d_{ij}}{r})^q}}
                             {\sum_{j=1}^{N} e^{-(\frac{d_{ij}}{r})^q}} \;.
\end{equation}
Structural components of a point set are characterized by the calculated
value of $\alpha$.
For example, points in a point-like structure have $\alpha \approx 0$
and pixels forming line-like structures have $\alpha \approx 1$.
Area-like structures
are characterized by $\alpha \approx 2$ of the pixels belonging to them.
A uniform distribution of points yields $\alpha \approx d$ which is equal
to the dimension of the configuration space.
The scaling indices for the point set form the frequency
distribution $N(\alpha)$
\begin{equation}
  N(\alpha) d\alpha = \#(\alpha \in [ \alpha,\alpha+d\alpha [)
\end{equation}
or equivalently the probability distribution
\begin{equation}
  P(\alpha) d\alpha = Prob(\alpha \in [ \alpha,\alpha+d\alpha [)
\end{equation}
The $P(\alpha)$-representation of a point set can be regarded as a
structural decomposition of the data where
the points are differentiated according to the local
morphological features of the structure elements to
which they belong to.

\subsection{Scaling Indices and Wavelets}

\begin{figure}
\centering
 \includegraphics[scale =0.345,angle=0]{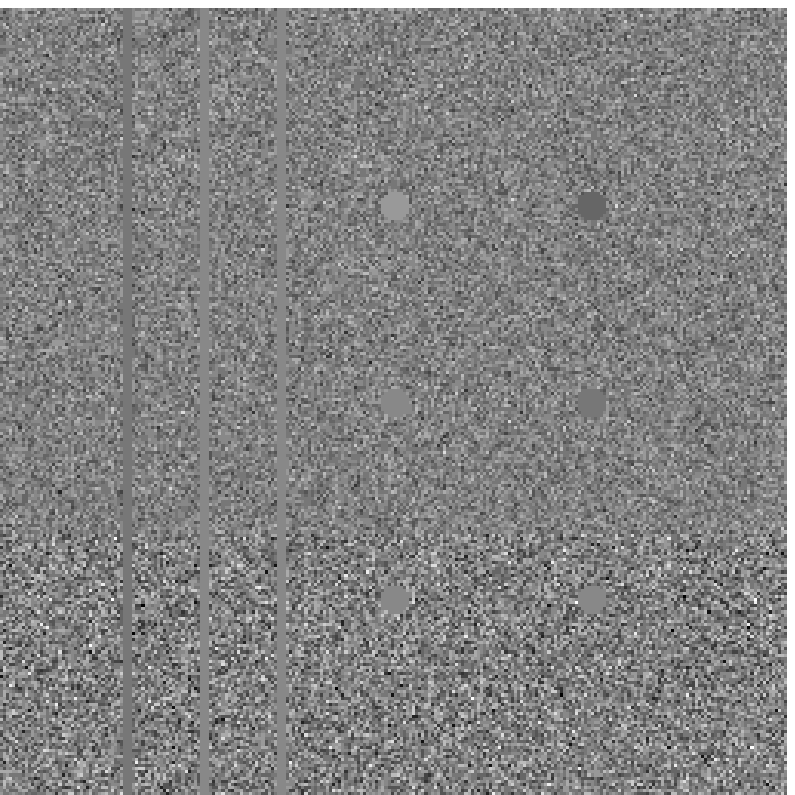}
  \includegraphics[scale =0.345,angle=0]{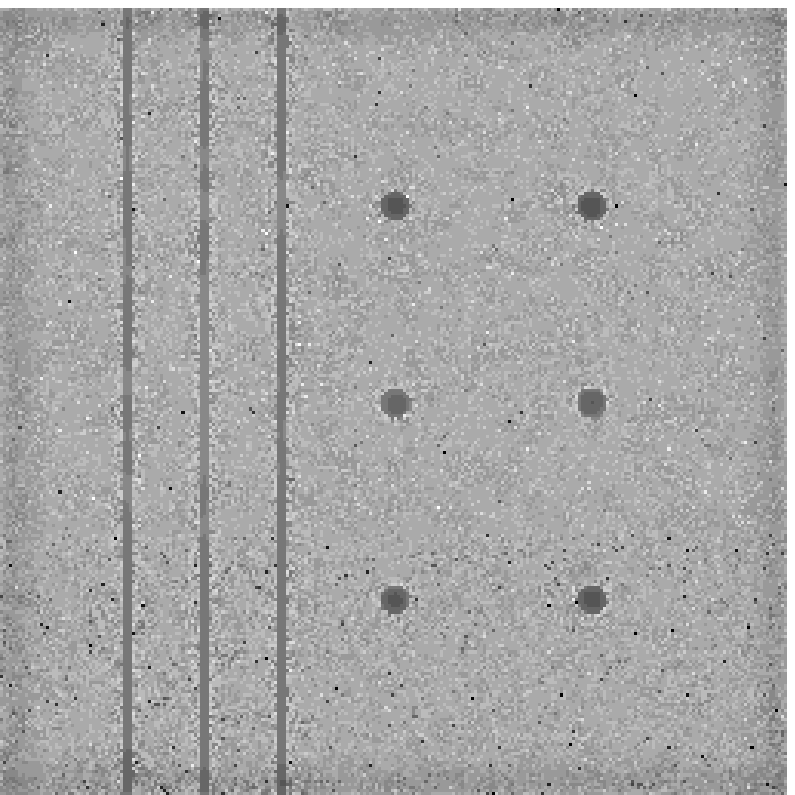}
  \includegraphics[scale =0.345,angle=0]{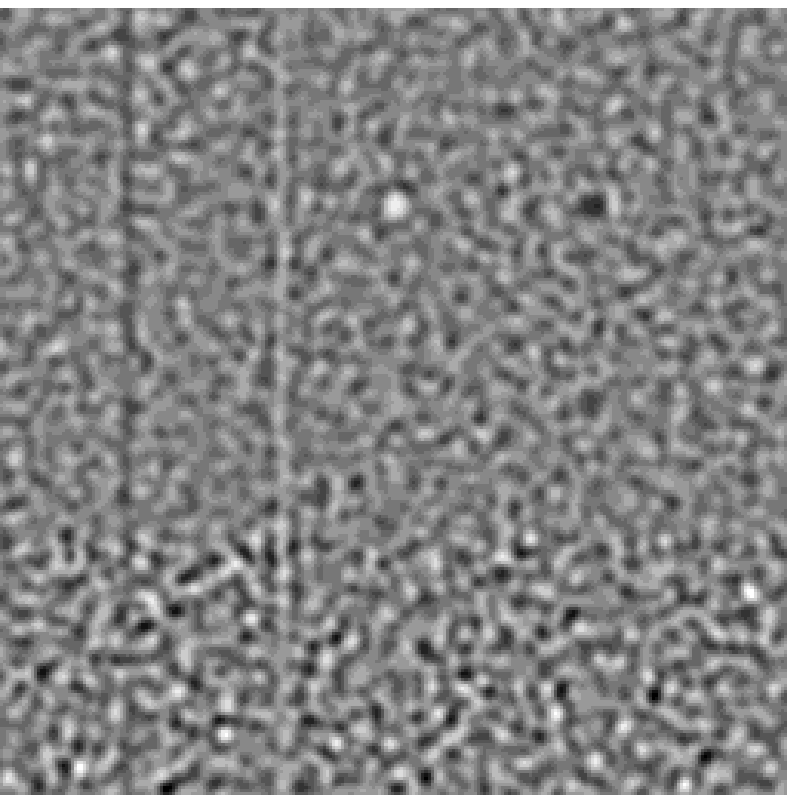} 
\caption{Left to right: Synthetic test image, response image of the scaling indices and response 
             image of the wavelet coefficients. The response images were normalised to the range of values
             of the scaling indices and wavelets respectively.  \label{fig2}}
\end{figure}

Some test statistics, with which non-Gaussian signatures have been detected
in the first and three-year WMAP-data, are based on a wavelet analysis of the 
CMB data, e.g. \citep{vielva04a, mcewen05a}. 
Beside the fact that wavelet-based tests yield one of the highest significances for non-Gaussianities 
in the WMAP-data, wavelets and scaling indices can also be considered to be similar to 
each other.  Both measures perform a {\it local} analysis 
of e.g. image data and they are calculated 
for different scales, which yields information about characteristic sizes of detected features.
On the other hand, there are obvious differences. While the scaling indices are a nonlinear
and non-bijective local filter, the wavelet transformation is a linear 
operation, namely  a scale- and space-dependent filtering of the image data with 
a wavelet function $\Psi$:
\begin{equation}
  w(b_x,b_y,r) = \int{dx dy  I(x,y) \Psi(x,y;b_x,b_y,r)} 
 \end{equation}
For the special case of spherical mexican hat wavelets  (SMHW), 
the 'mother wavelet' $\psi(d,r)$ is defined by:
 \begin{eqnarray}
  \Psi(d,r) & \equiv &\Psi(d,b_x=0,b_y=0,r)  \nonumber  \\
               & = & \frac{1}{(2\pi)^{1/2}r}  \left[ 2â - \left( \frac{d}{r} \right)^2 \right] e^{-d^2/2r^2} 
 \end{eqnarray}
 with $d=\sqrt{x^2+ây^2}$. 
It has been shown \citep{martinez02a} that SMHW yield superior results in detecting non-Gaussian 
signatures compared to other filter functions, e.g. Haar wavelets.\\
To consider the performance of the scaling indices in the context of known test statistics for 
non-Gaussianities we calculate and compare scaling indices and SMHW for two examples.\\
In the first simple example we generated a synthetic image which consists of a (white Gaussian) noisy background,
where the noise level was chosen to be twice as high in the lower third of the image than in the rest of the image. 
We interspersed three single-coloured lines and six disc-like elements  in the noisy background. The intensities of the
structural elements ranged from $0 \sigma$ to   $\pm 1 \sigma$ with respect to the standard deviations of the different
noisy backgrounds (see Figure \ref{fig2}). 

%%%%%%%%%%%%%%%%%%%%%%%%%%%%%%%%
%%%%%%%%%%%%%%%%%%%%%%%%%%%%%%%%

\begin{figure}
\centering
  \includegraphics[scale =0.55,angle=0]{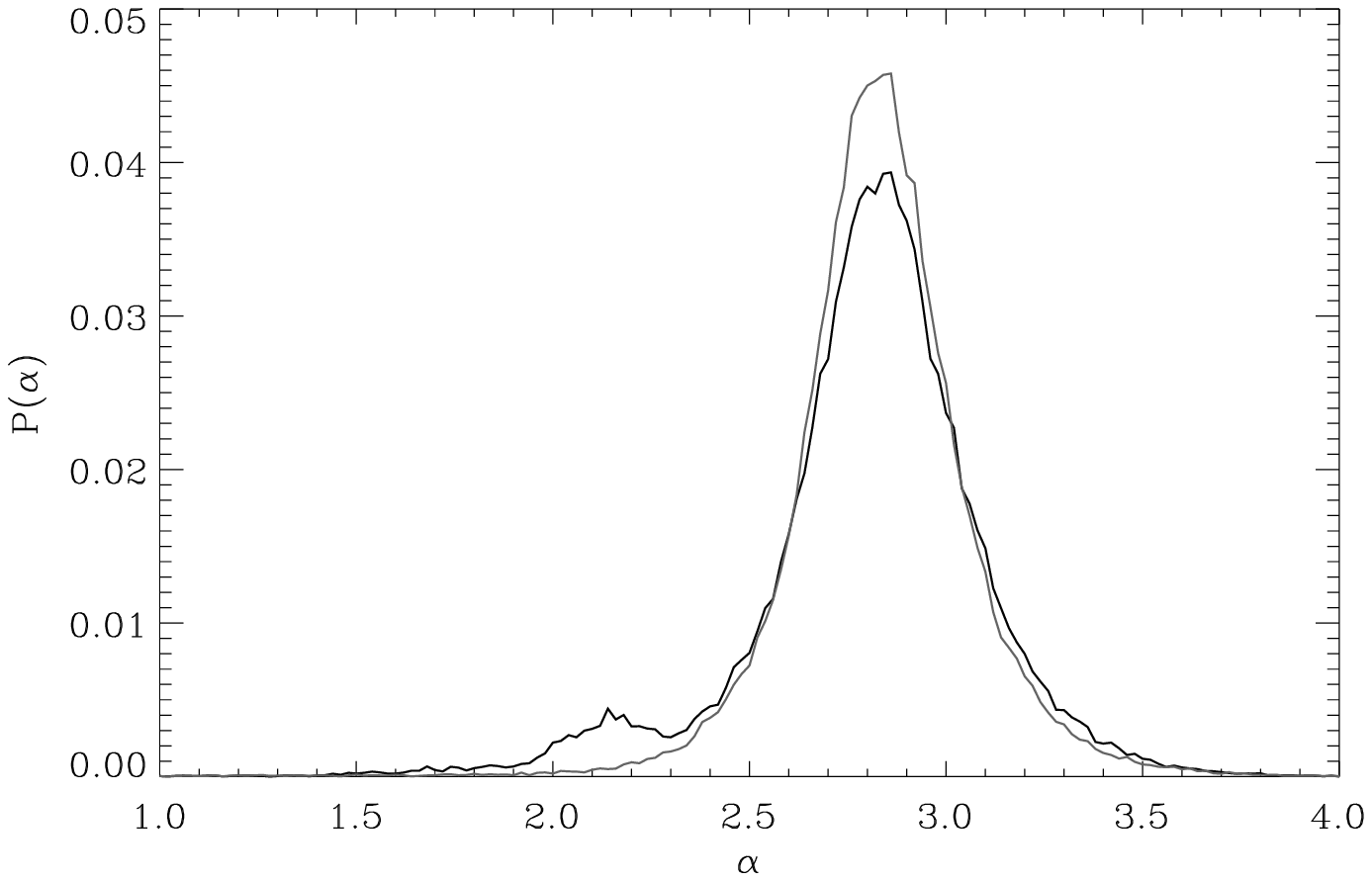}
  \includegraphics[scale =0.55,angle=0]{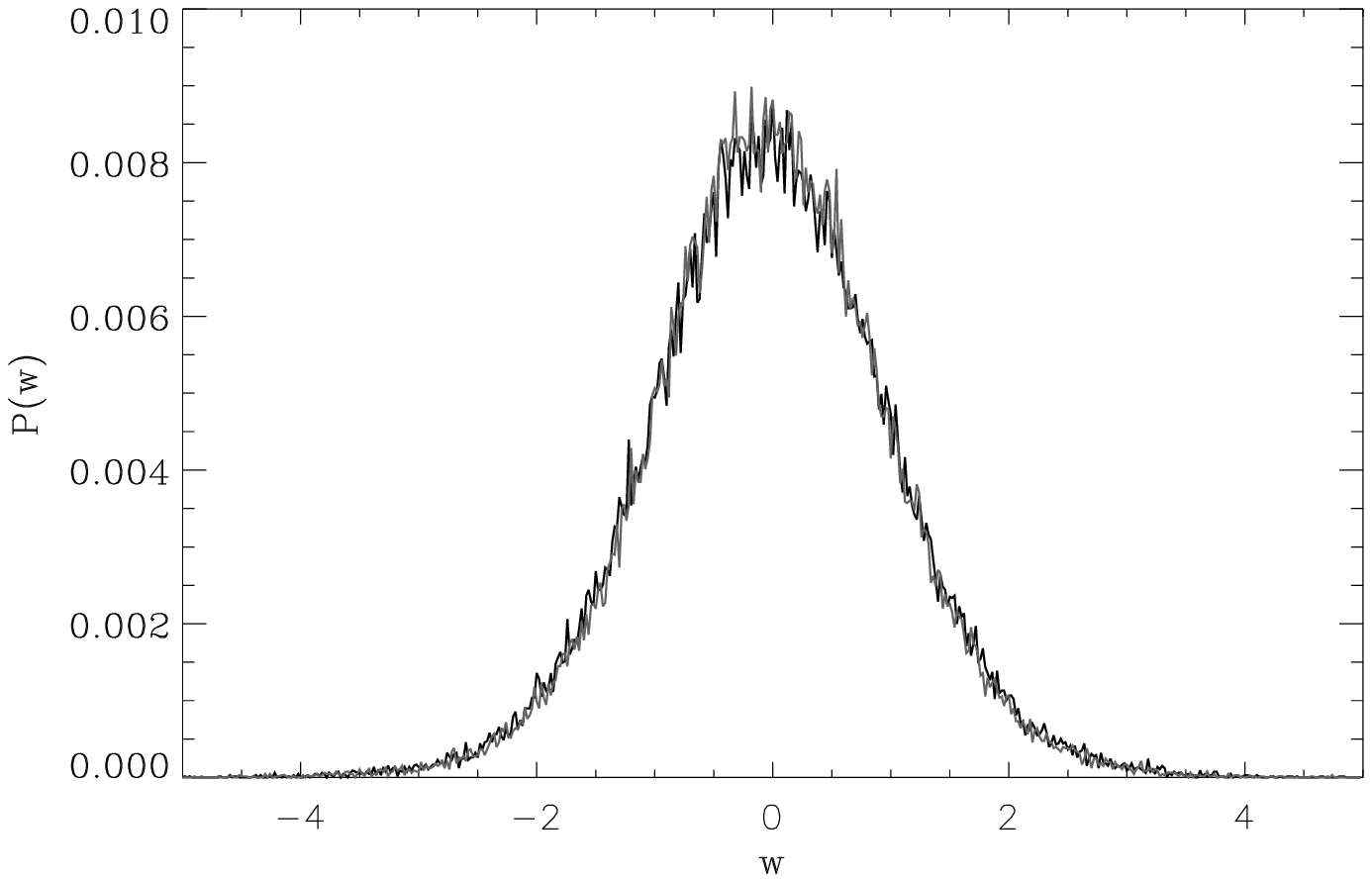} 
\caption{Probability distribution of the the scaling indices (above) and wavelet coefficients (below). The 
gray lines indicate the distributions for the pure noise image without the line- and disc-like structural elements.  \label{fig3}}
\end{figure}

We calculated the wavelets and scaling indices for a corresponding and well-suited scale. Figure  \ref{fig2} shows the grey-value 
coded response of the wavelet and scaling index filter. One can see that the wavelets can only detect those discs, 
which deviate most from the mean value of the noise. The lines are only hardly detected in the less noisy background.
The scaling indices, however, can detect all interspresed structural elements in the image, where
the (inner part of the) disc-like structures have lower $\alpha$-values than the three lines.
The observed differences in the images of the filter response find their reflections in the probability distributions $P(\alpha)$ and
$P(w)$ for the scaling indices and wavelet coefficients (Figure  \ref{fig3}).

%%%%%%%%%%%%%%%%%%%%%%%%%%%%%%%%
%%%%%%%%%%%%%%%%%%%%%%%%%%%%%%%%

\begin{figure}
\centering
  \includegraphics[scale =0.4,angle=0]{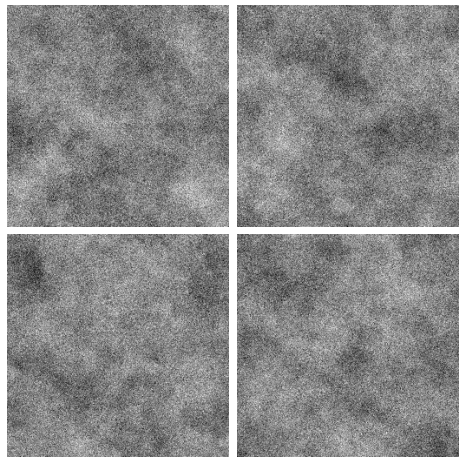} 
\caption{Simulated non-Gaussian CMB map with superimposed white additive noise (lower left) and three 
             surrogate realisations with the same power spectrum and amplitude distribution.  \label{fig4}}
\end{figure}

While the spectrum of scaling indices shows a clear second peak at $\alpha \approx 2.1$ and a larger tail 
towards lower $\alpha$-values, no visible deviations can be seen in the in the distribution of the wavelet coefficients 
when compared with those for a pure noisy image.  
With this example it becomes obvious that the wavelets are more sensitive to structures, which are associated with 
intensity variations of significant magnitude with respect to image noise.   
The scaling indices can, however, detect also structural features, which do not manifest themselves with significantly higher
(lower) intensity values but as intrinsic structural variations within the noise level.\\
In the second more realistic example we consider a realisation of CMB anisotropies due to the Kayser-Stebbin 
effect from cosmic strings on which white additive Gaussian noise was superimposed. The noise level is chosen with
rms signal-to-noise-ratio (S/N) of 0.5. 
For this image 20 surrogate maps were generated, which have the same power spectrum and amplitude distribution
as the original simulation (Figure  \ref{fig4}). 

%%%%%%%%%%%%%%%%%%%%%%%%%%%%%%%%

%%%%%%%%%%%%%%%%%%%%%%%%%%%%%%%%
\begin{figure}
\centering
  \includegraphics[scale =0.5,angle=0]{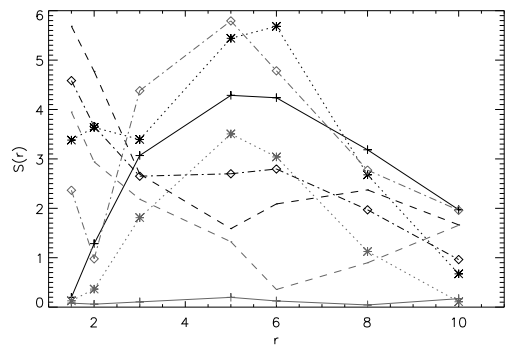} 
  \caption{Significances of the moments of the distribution of scaling indices (black) and wavelet coefficients (gray) 
               for the image in fig. \ref{fig4}.
              The lines and $+$ denote the mean, the dotteds line and $\star$ the standard deviation, 
              dashed lines the skewness and  dashdotted lines and $\diamond$ the kurtosis. \label{fig5}
              }
\end{figure}
%%%%%%%%%%%%%%%%%%%%%%%%%%%%%%%%

For more details about the simulation and its surrogates see \citet{raeth03a}.
We compare tests for non-Gaussianities based on wavelets and scaling indices.
In both cases we use the moments of the distributions of wavelet coefficient and scaling indices as test statistics,
namely the mean ($<\alpha>$), standard deviation ($\sigma_{\alpha}$), 
skewness ($\gamma_{\alpha}$) and kurtosis ($\kappa_{\alpha}$), which are defined by
 \begin{equation}
  <\alpha> = \frac{1}{N}\sum_{i=1}^{N} \alpha_i
 \end{equation}
 \begin{equation}
  \sigma_{\alpha} = \left( \frac{1}{N-1}\sum_{i=1}^{N} (\alpha_i- <\alpha>)^2 \right)^{1/2}
 \end{equation}
  \begin{equation}
  \gamma_{\alpha} =  \frac{1}{N}\sum_{i=1}^{N} \left( 
                     \frac{(\alpha_i- <\alpha>)}{\sigma_{\alpha} } \right)^3
 \end{equation}
  \begin{equation}
  \kappa_{\alpha} =  \frac{1}{N}\sum_{i=1}^{N} \left( 
                     \frac{(\alpha_i- <\alpha>)}{\sigma_{\alpha} } \right)^4 -3
 \end{equation}

 as test statistics. 
 For each statistic the degree of detected non-Gaussianity is measures in terms of significance $S$,
 \begin{equation}
  S= \left| \frac{M - <M>}{\sigma_M} \right|
 \end{equation}
 where $M$ is $ <\alpha>, \sigma_{\alpha}, \gamma_{\alpha}$ and 
 $\kappa_{\alpha} $ respectively.  
 The mean and standard deviation of the measure $M$ are 
 derived from the 20 surrogate reliasations.
 Figure  \ref{fig5} summarizes the results for the scaling indices and SMHW. Both classes of filters can well detect 
 the non-Gaussianities even at this high noise level. The highest significances occur at similar scales. 
 However, the best results were obtained for different moments. While for the scaling indices 
 the standard devaition performed best, we found for the wavelets the best discrimination with the kurtosis.   
The second best result for the scaling indices was obtained with the mean. 
Based on these findings we restrict ourselves to the analysis of the mean and standard deviation of the
distribution of scaling indices for the WMAP data.  
In summary we found in this example similar discrimination results for the two classes of filters, but for different
moments. The scaling indices gave - in general - better results for the lower moments of the distribution than the wavelets.
Since the lower moments are less sensitive to outliers of the distribution, one can arguably say that an analysis of 
the first moments of the scaling index distribution is 
statistically more stable and thus preferable.
   
\subsection{Weighted Scaling Indices for WMAP-Data}

To apply the SIM to the WMAP-data one has to find a proper representation
of the spherical data in a suitable embdding space.
Our aim in this work was to stay as close as possible to the original data, i.e. to maintain the 
spherical symmetric character of the data set  and thus to omit any projection onto a flat space. 
On the other hand, we strove to find a proper three-dimensional embedding of the data
according to the three free parameters of each pixel on the unit sphere, 
namely their angles and temperature.
For this, we chose the following approach:\\
Consider a temperature map $I(\theta,\phi)$, where
each pixel is assigned with a temperature value $I(\theta,\phi)$.
The position of the pixel on the (unit) sphere is given by the
two angles $\theta$ and $\phi$. In addition, it contains
the temperature information $I(\theta,\phi)$.
One possible three-dimensional representation of the WMAP-data,
in which both the spatial and temperature information of each pixel
is simultaneously taken into account, is given by
\begin{eqnarray}
x  & = & (R + dR) \cos(\phi)\sin(\theta)\\
y  & = & (R + dR) \sin(\phi)\sin(\theta)\\
z  & = & (R + dR) \cos(\theta)
\end{eqnarray}
with
\begin{equation}
dR = a (T(\theta,\phi) -<T>)/ \sigma_{T} \;.
\end{equation}

\begin{figure}
\centering
  \includegraphics[scale =0.25,angle=0]{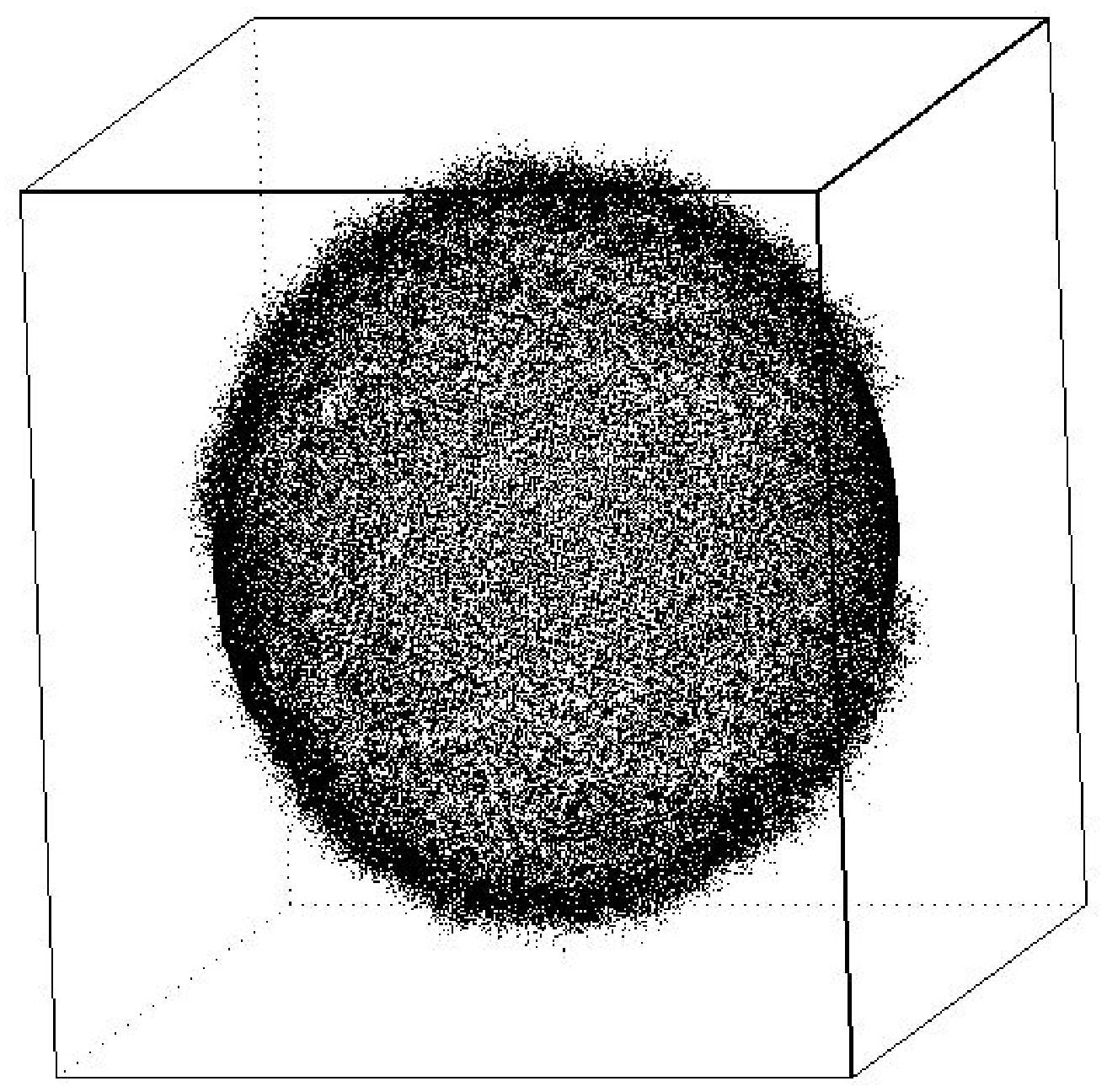}
  \includegraphics[scale =0.25,angle=0]{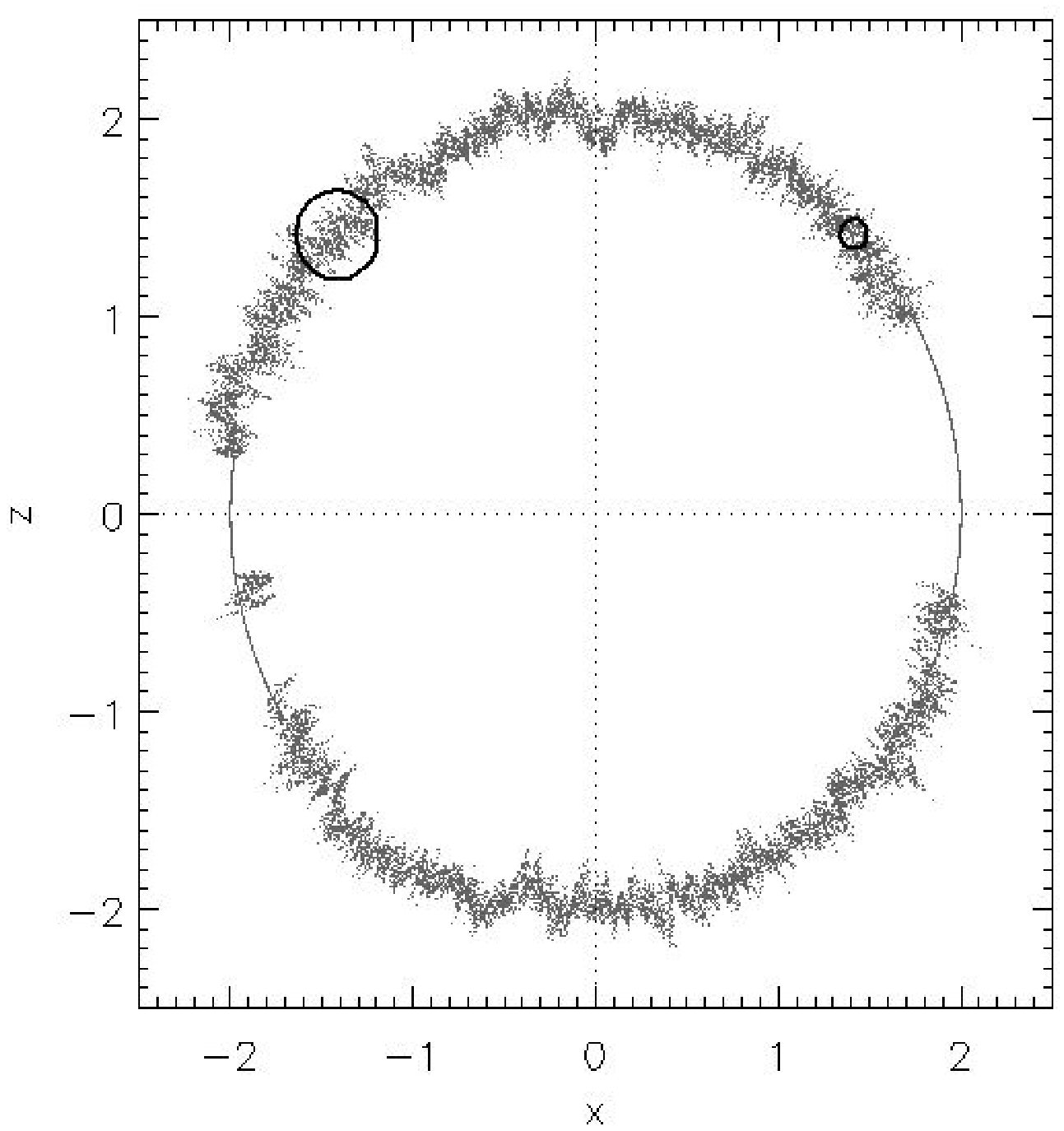}
     \includegraphics[scale =0.25,angle=0]{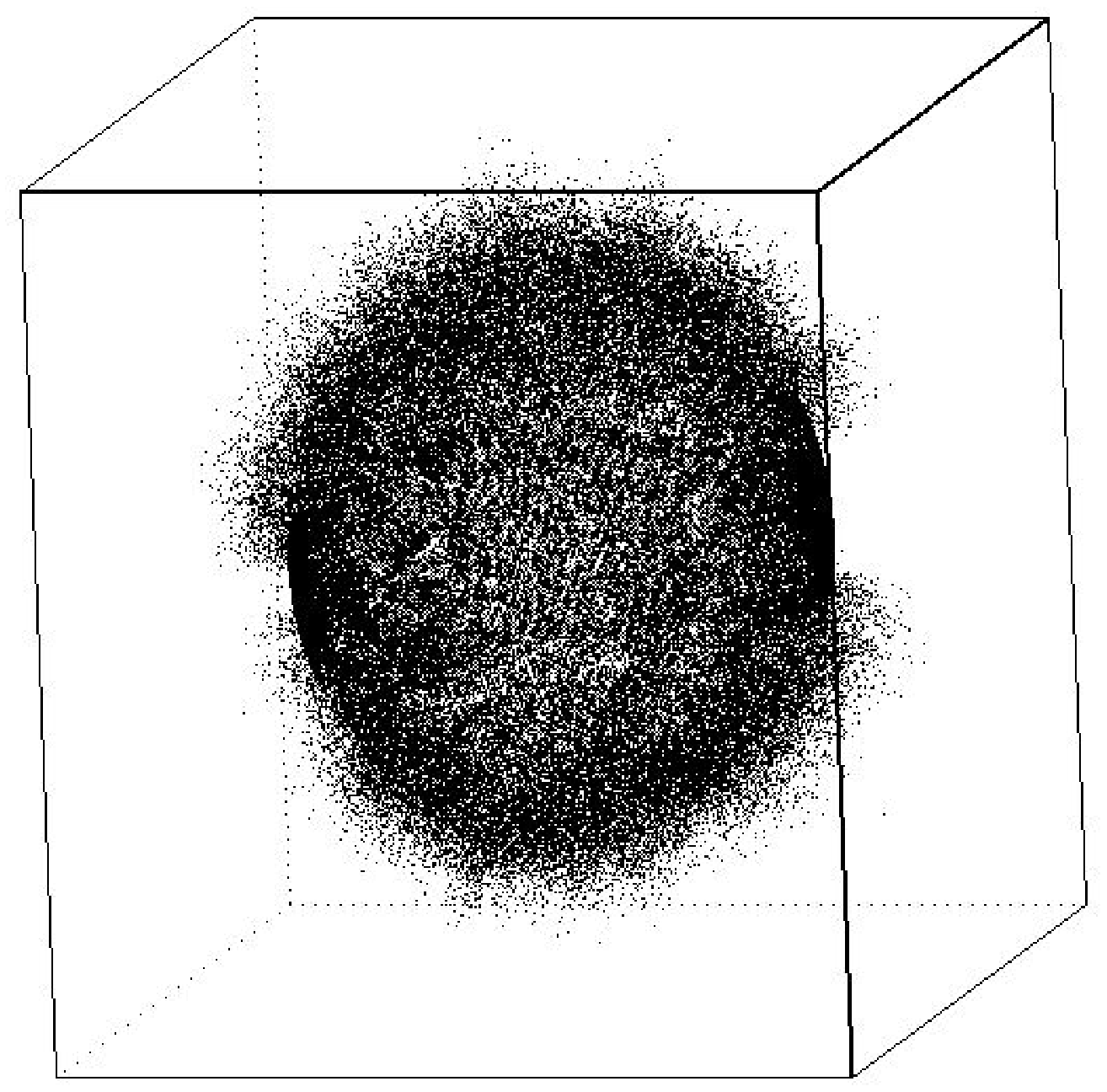}
  \includegraphics[scale =0.25,angle=0]{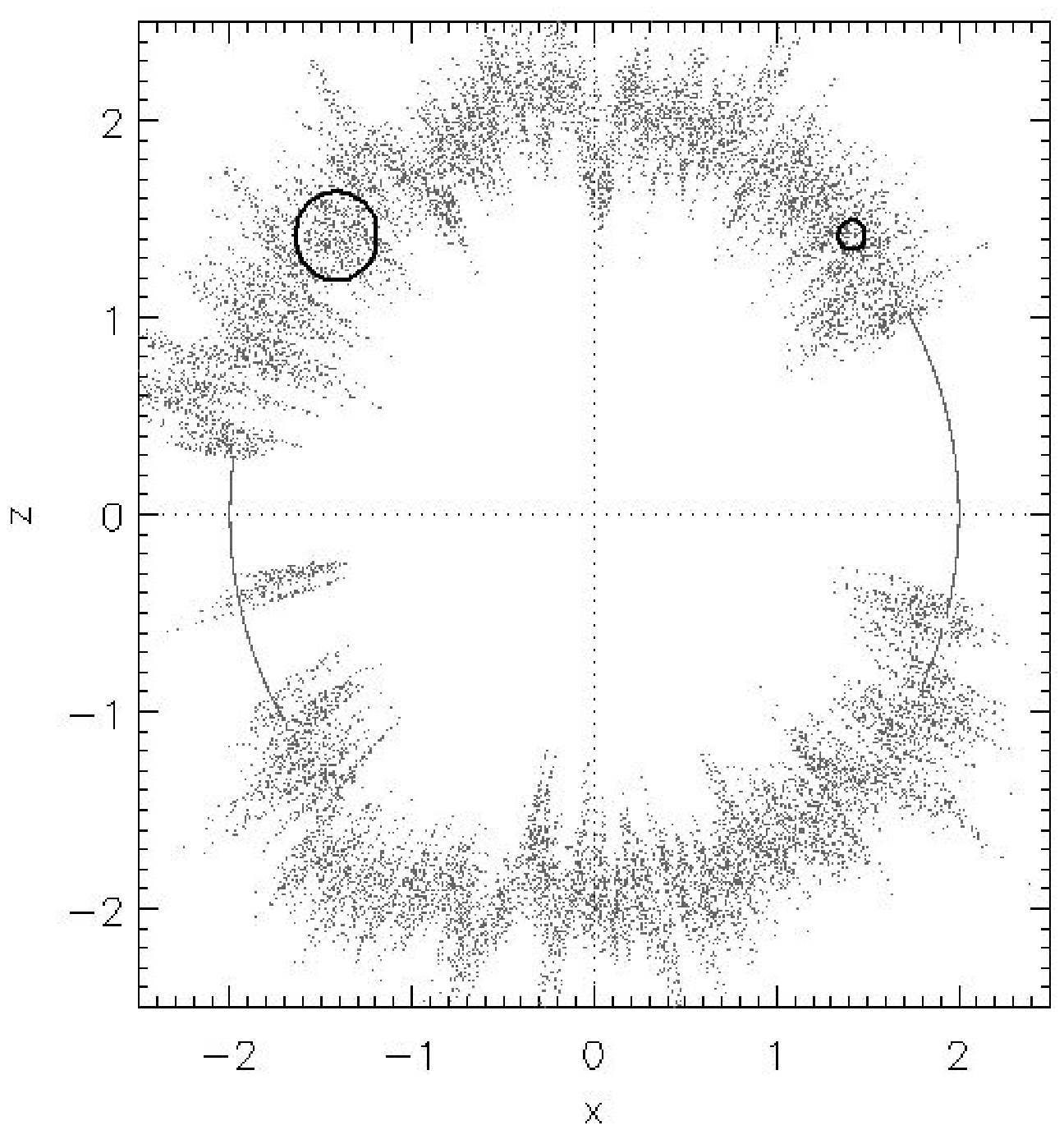}
   
  \caption{Left:  WMAP-data represented as a three-dimensional point distribution. Right:  $x,z$-projection 
               of all points with $ | y |<0.05$. Above:  $a=0.075$, below: $a=0.225$. The black circles indicate 
               the scaling ranges $r=0.075$ and $r=0.225$ respectively. A good sensitivity to the temperature 
               fluctuations at a given scale $r$ is obtained for $r=a$. \label{fig6}}
\end{figure}

By introducing the term $dR$ the temperature anisotropies are transformed 
to variations in the radial direction around the sphere.
The normalisation ensures that $dR$ has zero mean and a
standard deviation of $a$.
The parameters $R$ and $a$ are free (scale) parameters, which control the size 
of the temperature-induced radial jitter relative to its spatial extent.
These two parameters $R$ and $a$, together with the scaling range parameter $r$ for the calculation
of the scaling indices have to be properly set so that the scaling indices are sensitive to the 
temperature fluctuations in the chosen (spatial) scaling range (see Figure \ref{fig6}). 

%%%%%%%%%%%%%%%%%%%%%%%%%%%%%%%%
%%%%%%%%%%%%%%%%%%%%%%%%%%%%%%%%

To achieve this, we coupled the parameter $a$, which controls the width of the radial jitter with
the scaling range $r$ and set for each scaling range $a=r$.  
The size of the the sphere $R$ has to be chosen large enough with respect to the radial
jitter $dR$, so that the scaling indices remain a local measure and their calculation are
not affected by pixels on the opposite side of the sphere with a large $dR$. 
Hence, we deliberately set $R=2$ for all calculations presented below.
We performed some tests for the WMAP data and a subset of simulations using
different values of $R$, e.g. $R=1.5$, and found only marginal differences in the results.\\  
Figure  \ref{fig6} shows the representation of the WMAP data as
a point set $P=\{\vec{p_i}=(x_i,y_i,z_i)\}, i=1,\ldots,N_{pixel}$ and  two-dimensional projections 
of all data with $|y|<0.05$.
For each point $p_i$ a set of weighted scaling indices $\alpha$
for ten different values of $r$, $r=0.025, 0.05, 0.75,\ldots,0.25$
is calculated, which (roughly) corresponds an angular resolution $\psi$
of $\psi=1.4^{\circ}, 2.8^{\circ},\ldots, 14.3^{\circ}$.\\
Thus, the weighted scaling indices can also
be interpreted as a filter response of a local nonlinear
filter acting on the spherical CMB data.
Global quantities of the $\alpha$-distribution as well as the
probability density $P(\alpha)$ are used to derive statistics quantifying
the structures in the WMAP-data and respective simulations.

\section{Results}

%%%%%%%%%%%%%%%%%%%%%%%%%%%%%%%%
\begin{figure}
\centering
 \includegraphics[scale =0.5,angle=0]{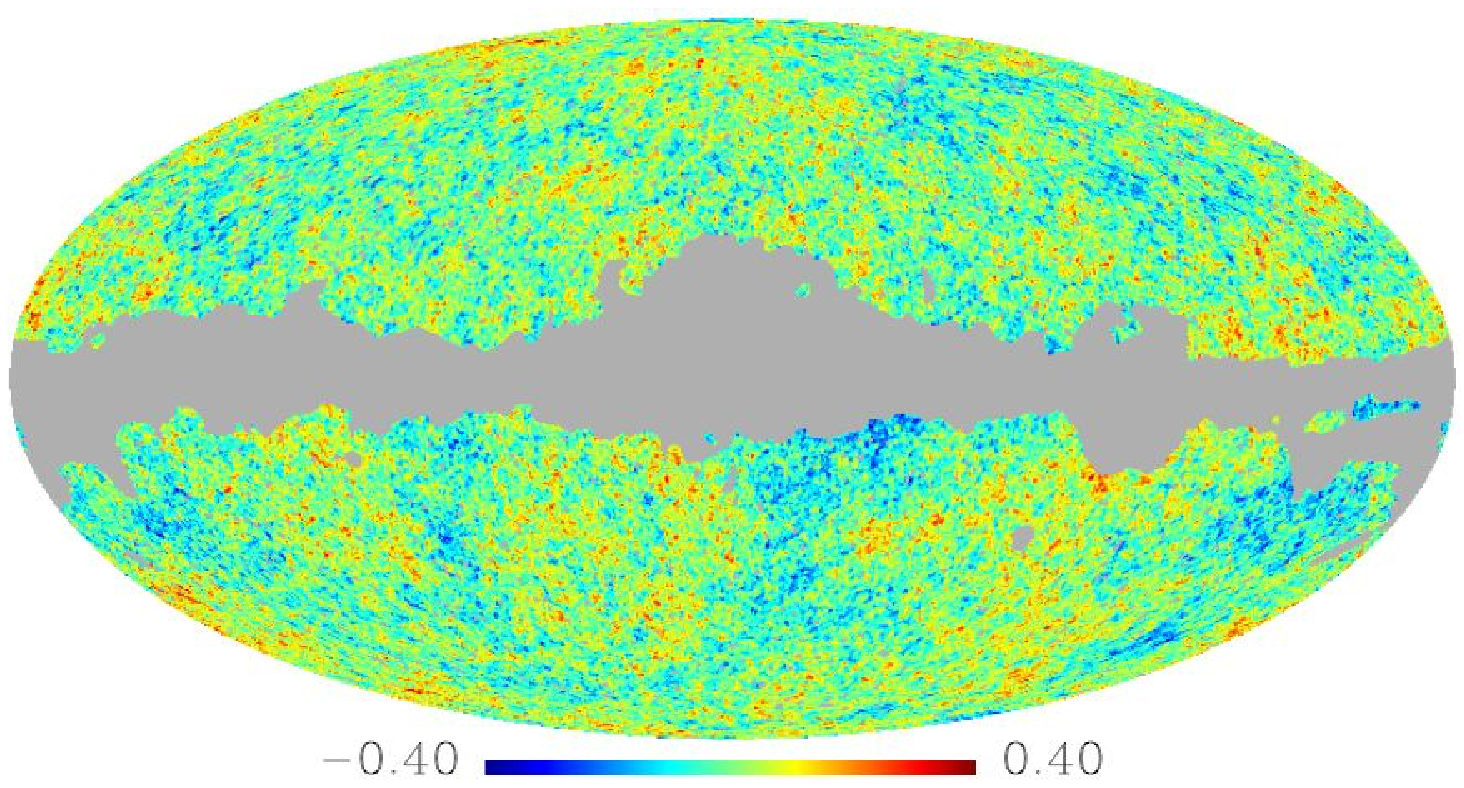}
  \includegraphics[scale =0.5,angle=0]{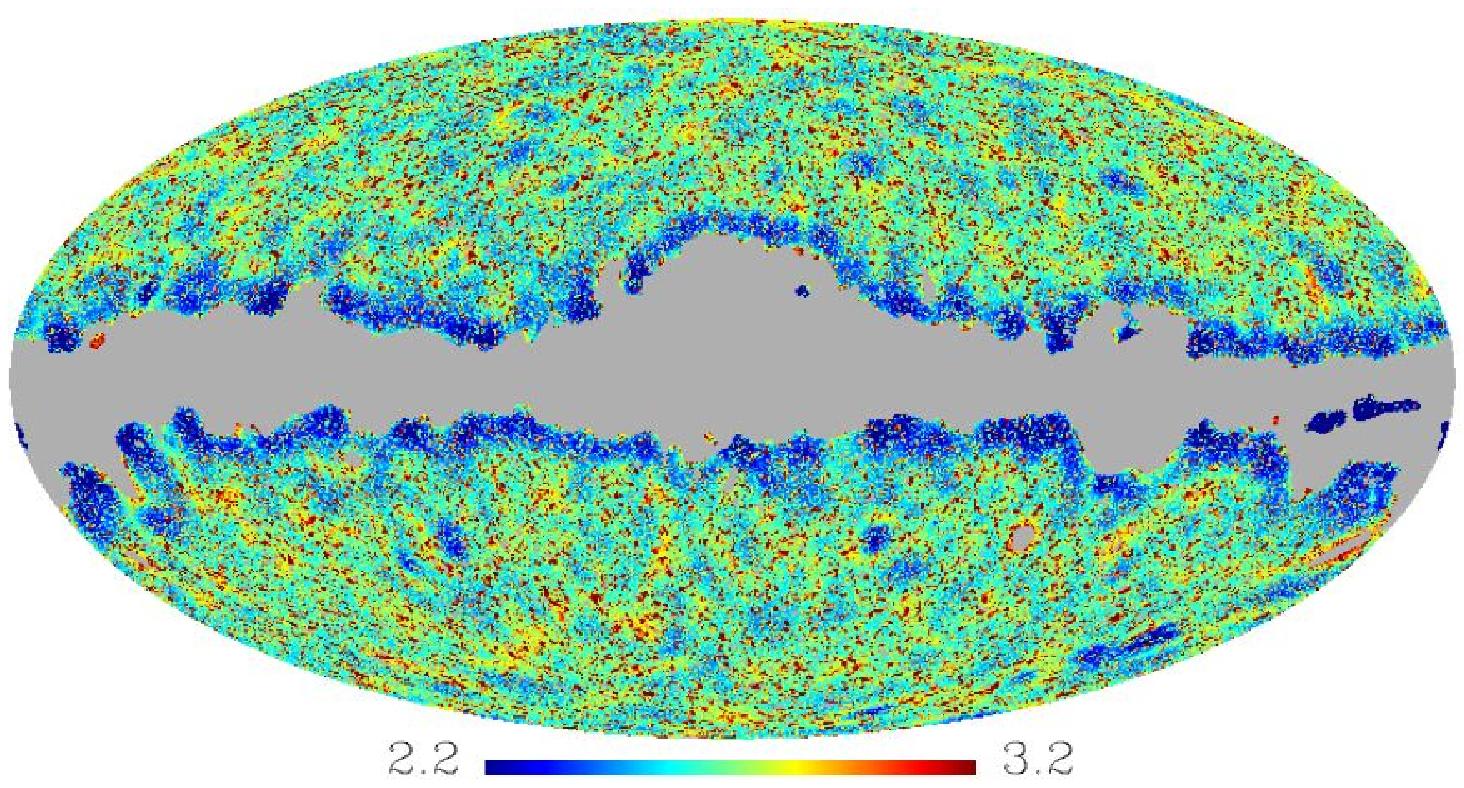}
  \includegraphics[scale =0.5,angle=0]{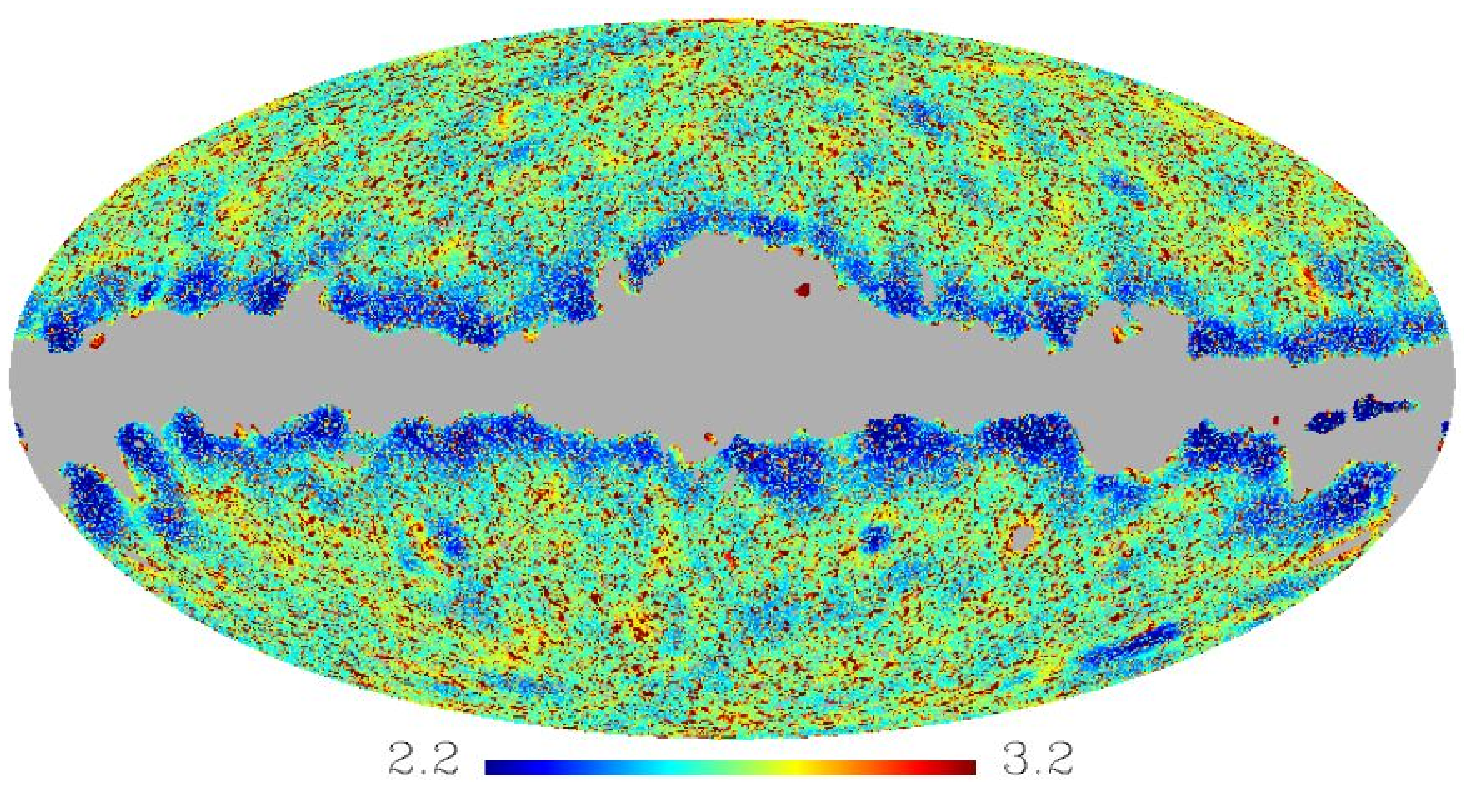}
  \includegraphics[scale =0.5,angle=0]{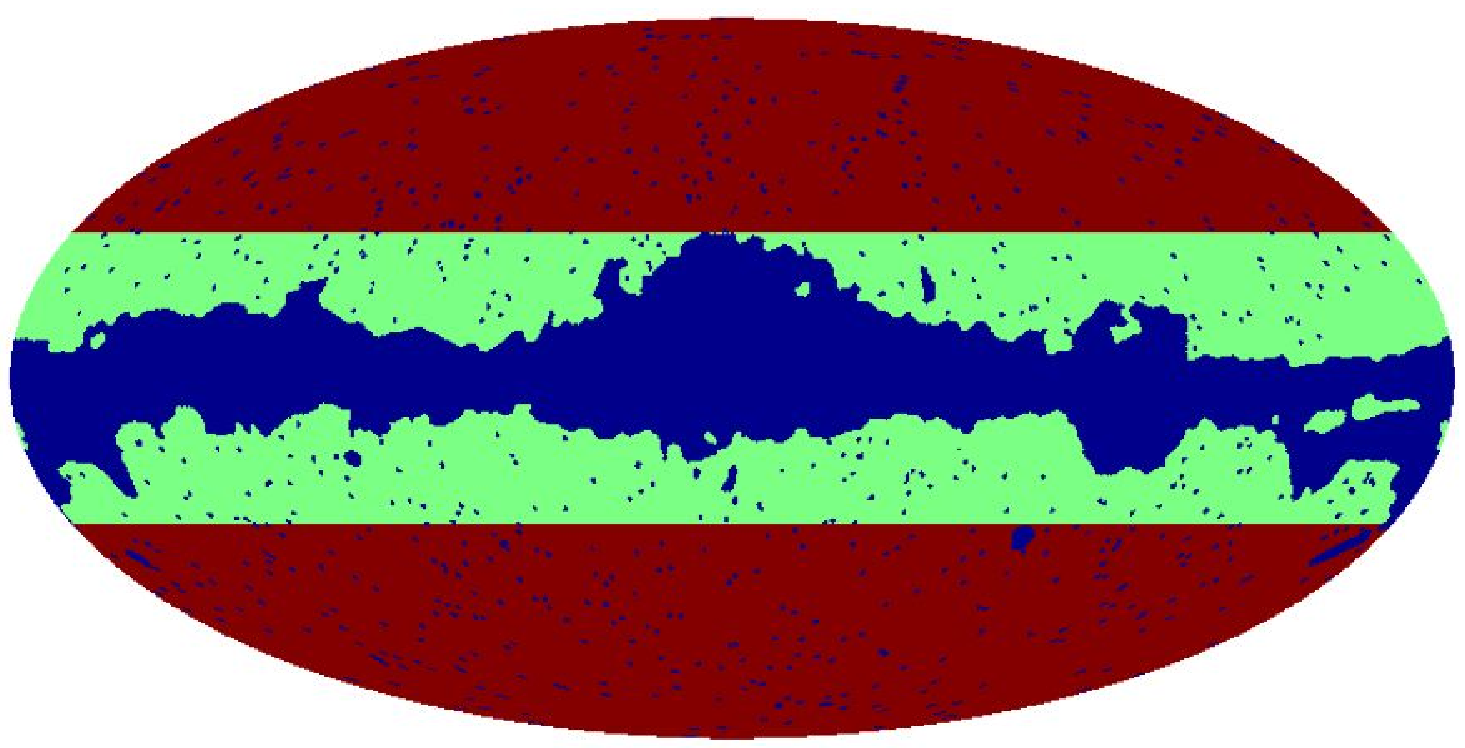}
\caption{From top to bottom: Co-added WMAP-map with Kp0-mask,
              color-coded $\alpha$-response for $r=0.175$ and $r=0.225$ 
              and  Kp0- and extended mask. \label{fig7}}
\end{figure}
%%%%%%%%%%%%%%%%%%%%%%%%%%%%%%%%

In Figure  \ref{fig7} the preprocessed foreground-cleaned coadded map of the WMAP data and 
the $\alpha$- response for two different radii $r$ is shown. It is obvious that for these larger 
scales the pixels in the vicinity of the Kp0-mask are affected by boundaries effects, which 
lead to systematically lower values for $\alpha$. 
This effect, which is only due to the lack of 
data points within the Kp0 mask, is the same for both 
the WMAP-data and the simulations, so that global measures based on the scaling 
indices, e.g. the moments of the probability distribution, are systematically affected in the same way.
However, some smaller local effects may be diluted  by the boundary effects. 
In order to have a cleaner map we also considered only scaling indices in a more conservative  mask, where only pixels
 with $ \left| b \right| > 30$ ($b$: Galactic latitude) were selected (Figure \ref{fig7}). Note that for 
the calculation of the scaling indices all pixels outside the Kp0-mask were taken into account.
In the following we always explicitly specify to which chosen 
mask the presented results refer.
%%%%%%%%%%%%%%%%%%%%%%%%%%%%%%%%
\begin{figure}
\centering
 \includegraphics[scale =0.5,angle=0]{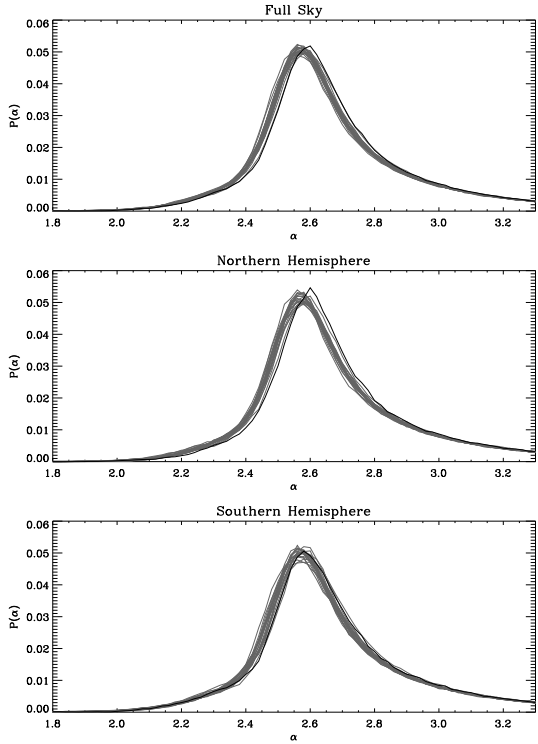}
  \caption{Probability density $P(\alpha)$ for the co-added WMAP
           data and $20$ simulations for $r=0.175$ and the Kp0-mask. 
             \label{fig8a}}
\end{figure}
%%%%%%%%%%%%%%%%%%%%%%%%%%%%%%%%

%%%%%%%%%%%%%%%%%%%%%%%%%%%%%%%%
\begin{figure}
\centering
  \includegraphics[scale =0.5,angle=0]{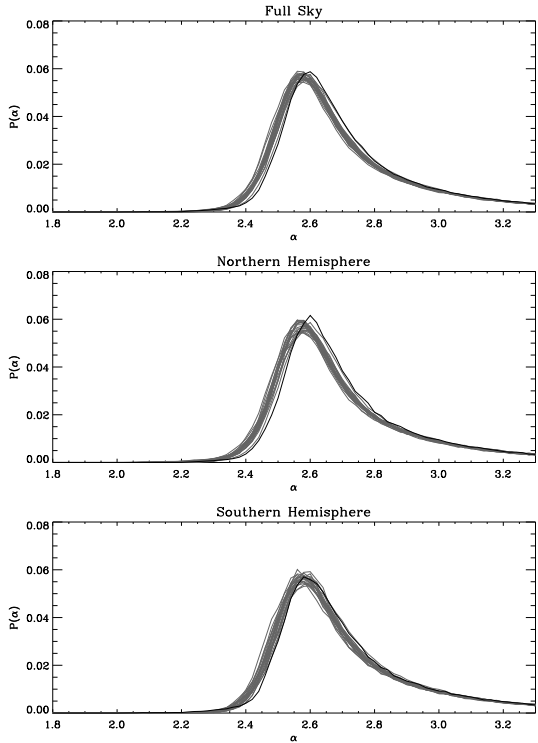}
  \caption{Same as Figure \ref{fig8a} but for the extended mask. \label{fig8b}}
\end{figure}
%%%%%%%%%%%%%%%%%%%%%%%%%%%%%%%%

The probability densities $P(\alpha)$ of the scaling indices 
for one selected scale ($r=0.175$) and both masks 
are displayed in the Figures  \ref{fig8a} and \ref{fig8b} for the WMAP-data and a subset of 20 simulations. 
One immediately realizes that for both masks the probability density for 
the WMAP data is shifted towards  higher values, 
which indicates that the underlying temperature fluctuations for the
observed data resemble more 'unstructured', i.e. purely random and uniform fluctuations 
than the simulations. 
This effect is more pronounced in the northern hemisphere than in 
the southern. Furthermore, the distributions are broader for the simulations than for the 
WMAP-data, indicating that the simulations exhibit a larger structural variability than the 
observed data. 
%%%%%%%%%%%%%%%%%%%%%%%%%%%%%%%%
\begin{figure}
\centering
 \includegraphics[scale =0.5,angle=0]{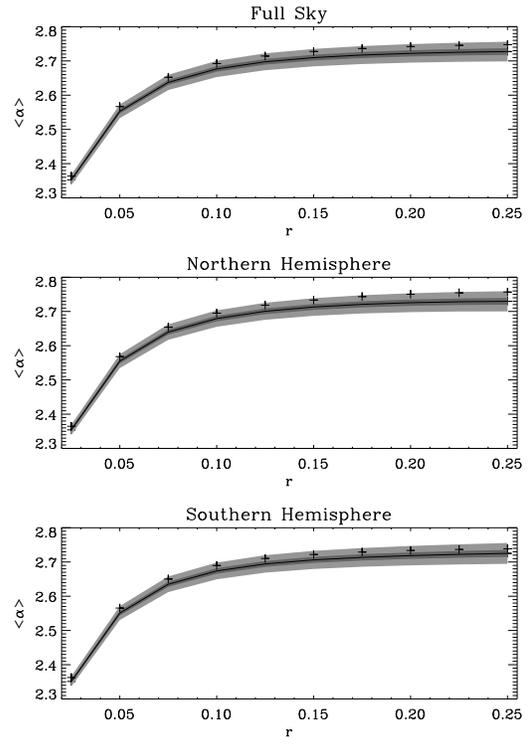}
 \caption{Scaling index mean statistics as a function of the scaling range $r$ for the full sky (above), 
              northern (middle) and southern hemisphere (below) using the Kp0-mask.  $+$ denote
               the values for WMAP, the dark gray and light gray areas indicate the $1\sigma$ and $3\sigma$ 
               regions around the mean value (black line) as derived from the simulations.  \label{fig9}}
\end{figure}
%%%%%%%%%%%%%%%%%%%%%%%%%%%%%%%%
\begin{figure}
\centering
 \includegraphics[scale =0.5,angle=0]{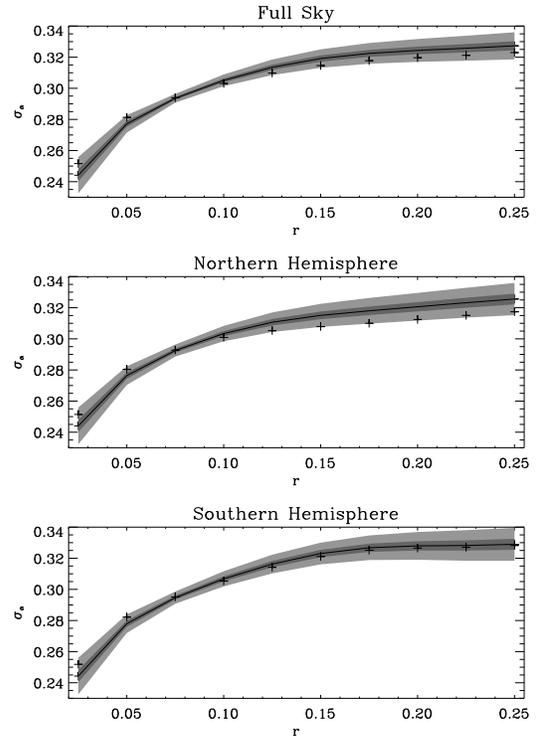}
 \caption{Same as fig. \ref{fig9} but for the standard deviation $\sigma_{\alpha}$. \label{fig10}}
\end{figure}

\begin{figure}
\centering
 \includegraphics[scale =0.5,angle=0]{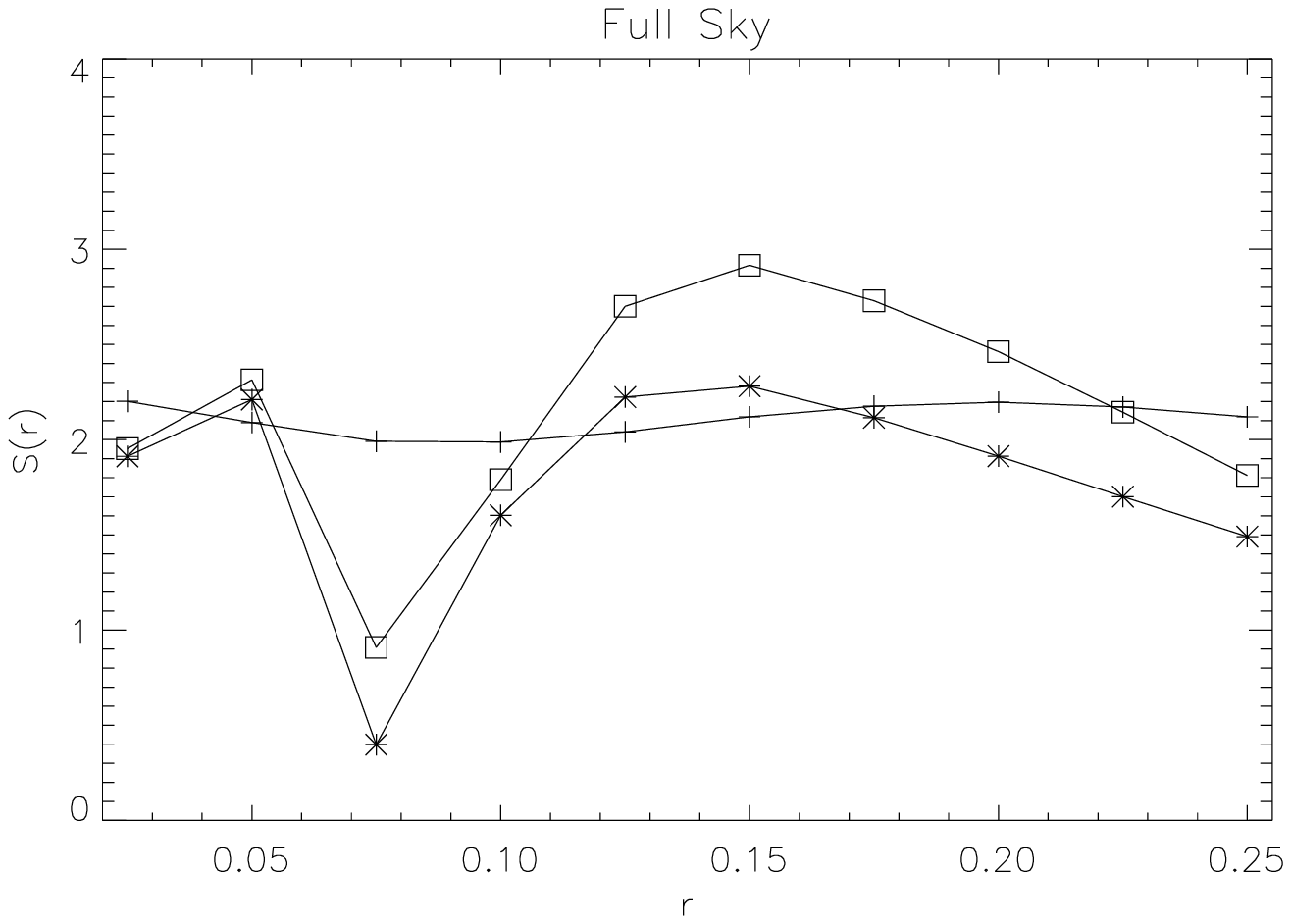}
  \includegraphics[scale =0.5,angle=0]{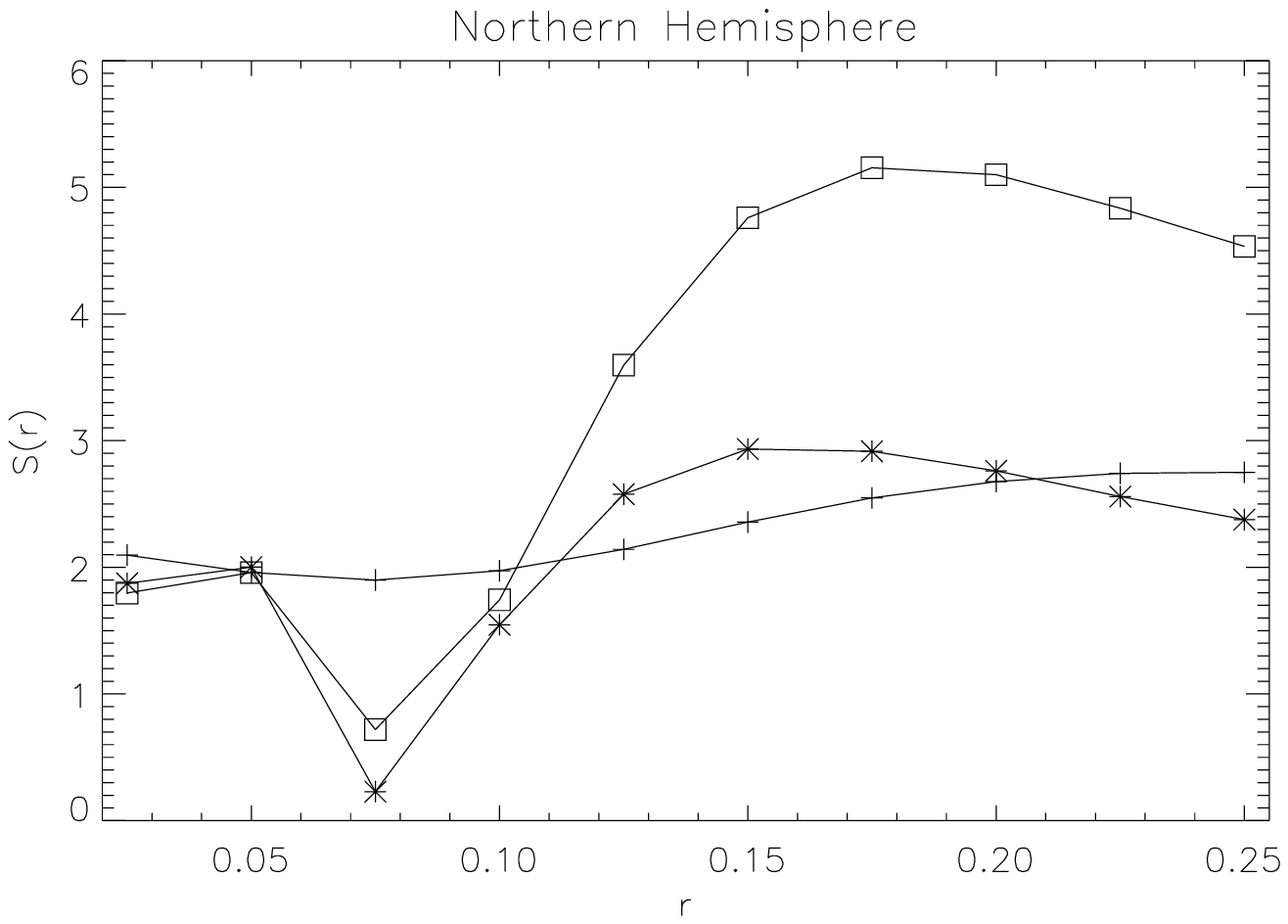}
  \includegraphics[scale =0.5,angle=0]{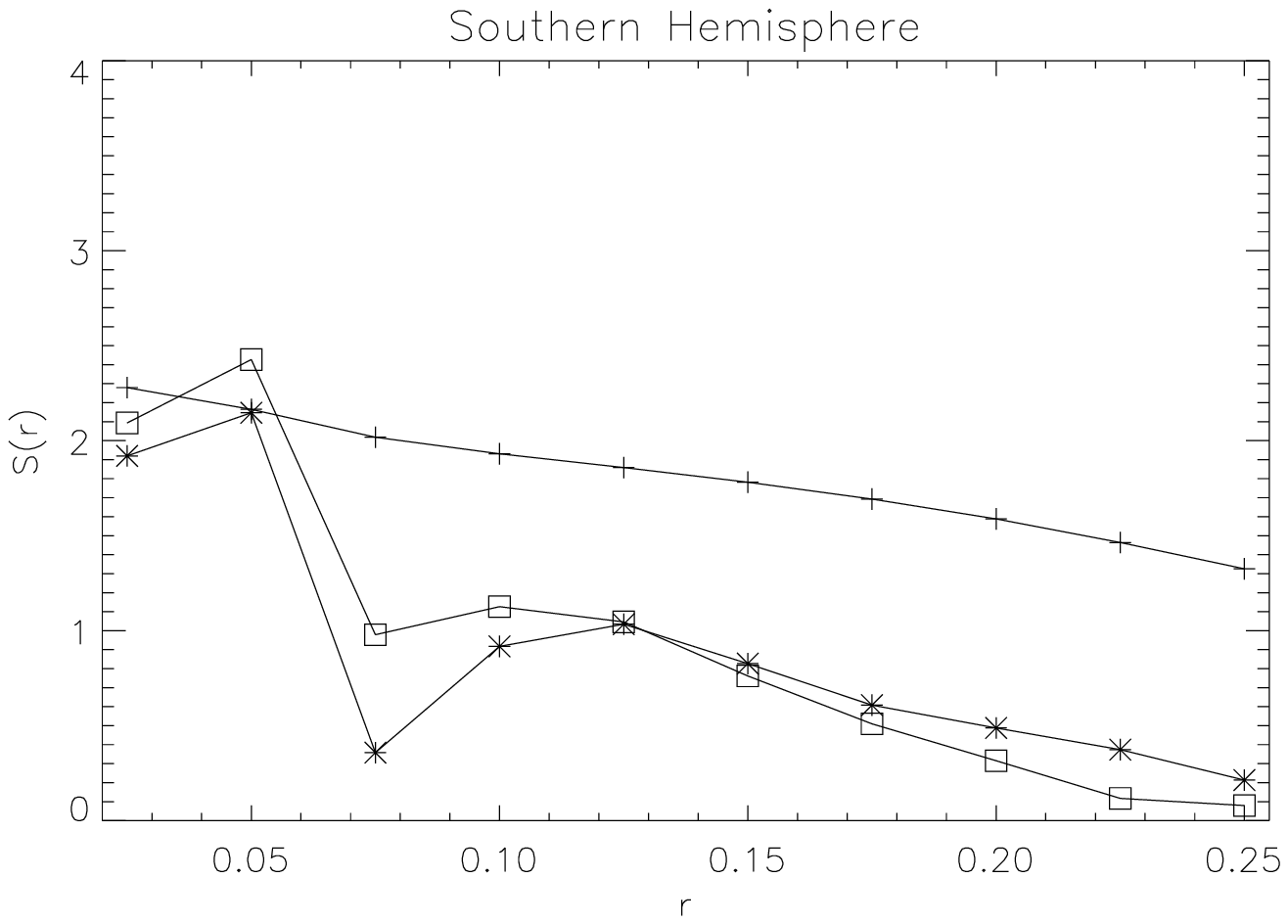}
\caption{Significances of the (combined) moments of the $\alpha$-distribution of the WMAP-data with the Kp0-mask 
              as a function of the scaling range $r$.
              $+$ denotes the mean, $\star$ the standard deviation and 
              the boxes $\chi^2$-combination of the mean and standard deviation. 
                            \label{fig11}}
\end{figure}

\begin{figure}
\centering
 \includegraphics[scale =0.5,angle=0]{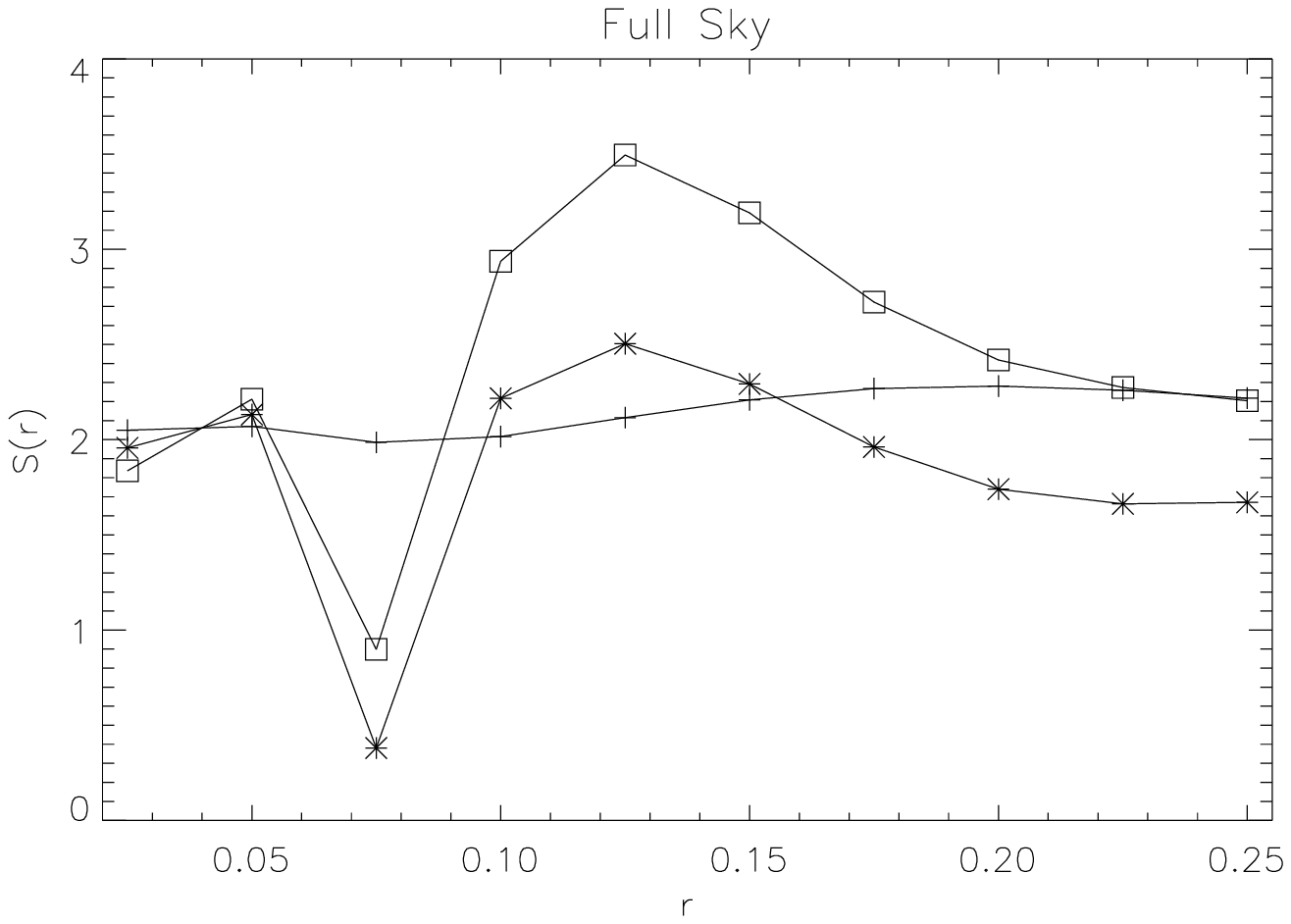}
  \includegraphics[scale =0.5,angle=0]{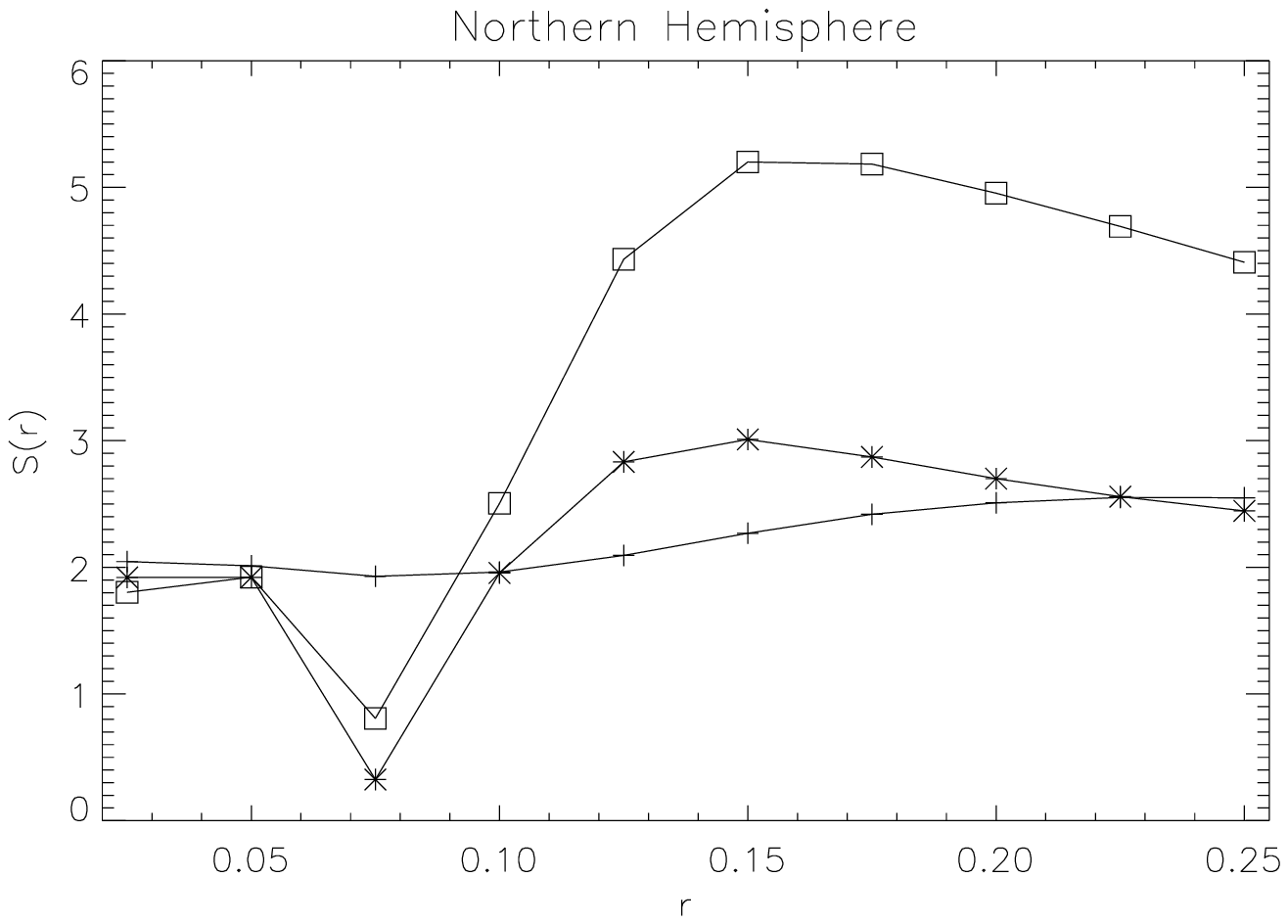}
  \includegraphics[scale =0.5,angle=0]{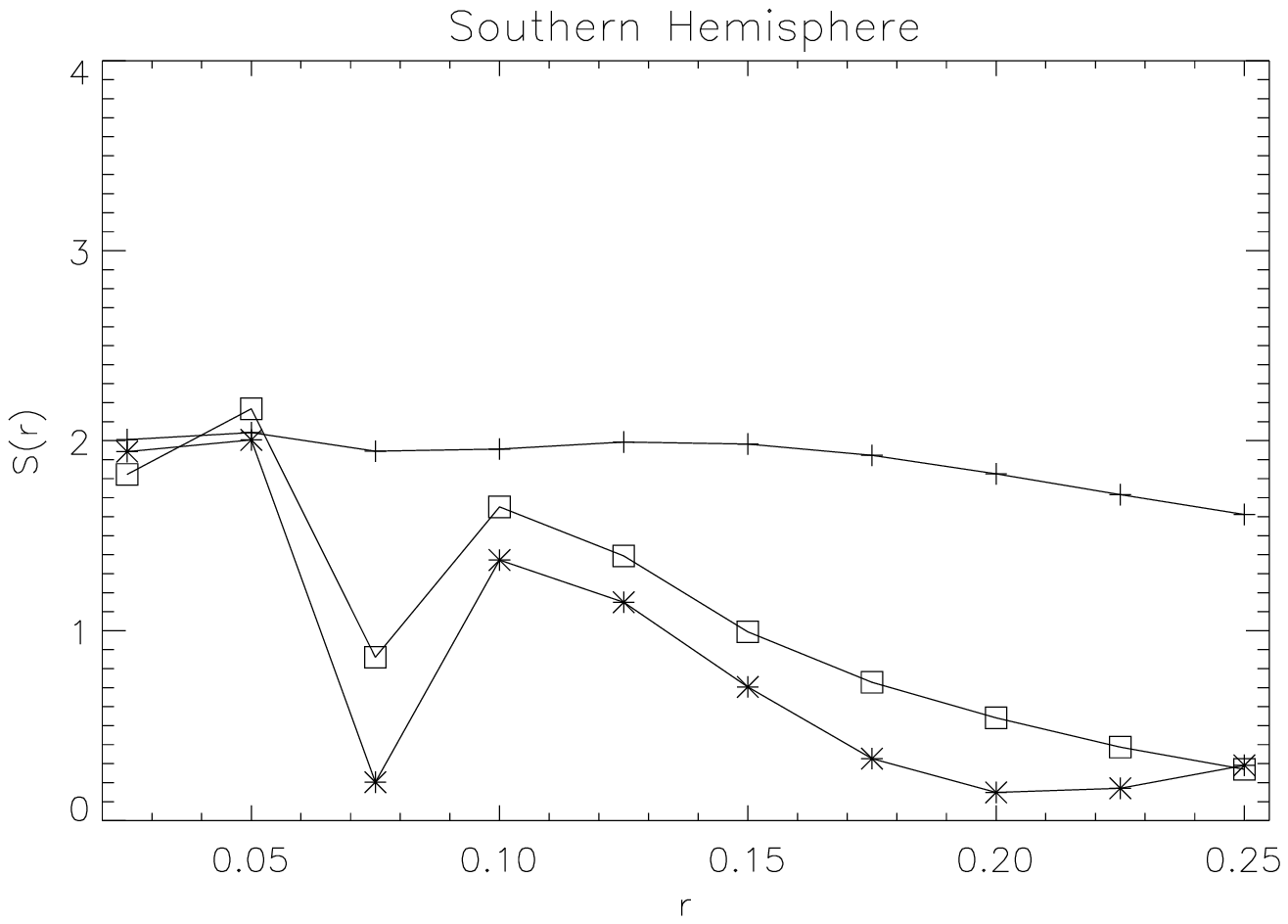}
\caption{Same as Fig. \ref{fig11} but for the extended mask. \label{fig12}}
\end{figure}

%%%%%%%%%%%%%%%%%%%%%%%%%%%%%%%%
These effects can more rigorously be quantified by calculating the mean and  
standard deviation for the distribution of scaling indices as calculated for different 
scaling ranges. Figures  \ref{fig9} and  \ref{fig10} show these results. 
For scales larger than $r=0.1$ the mean of the 
scaling indices is always systematically higher for the WMAP 
than for the simulations. The effect is much more pronounced in the northern hemisphere. 
For the standard deviation we  observe for the same scales significantly lower values for WMAP 
in the northern hemisphere  and slightly higher ones for the southern sky. 
For the full sky these two effects cancel each other so that the observed deviations 
to lower values are no longer so significant.\\â
Beside the mean and standard deviation we additionally considered a combination 
of these two test statistics, namely a diagonal $\chi^2$-statistic
\begin{equation}
\chi^2=  \sum_{i=1}^2  \left[ \frac{M_i - <M_i>}{\sigma_{M_i}} \right]^2 \;,
\end{equation}
where $M_1=<\alpha>$ and  $M_2=\sigma_{\alpha}$.
This statistic is computed for both the simulations and the observed moments.
Note that we follow the reasoning of \citet{eriksen04a} and choose 
a diagonal $\chi^2$-statistics, because also in our case the moments are highly correlated 
leading to high values in the off-diagonal elements of the cross-correlation matrix. 
But if the chosen model is a proper  description of the data, {\it any} combination of 
measures should yield statistically the same  values for the observations and the simulations. 
The significances of the deviations of the WMAP-data from the simulations
\begin{equation}
S = \left| \frac{M - <M>}{\sigma_M} \right|
\end{equation}
($M=<\alpha>, \sigma_{\alpha}$ and $\chi^2$)
are shown in Figures  \ref{fig11} and  \ref{fig12} for the Kp0-mask and the extended mask.

%%%%%%%%%%%%%%%%%%%%%%%%%%%%%%%%

%%%%%%%%%%%%%%%%%%%%%%%%%%%%%%%%
%%%%%%%%%%%%%%%%%%%%%%%%%%%%%%%%

The mean  $<M>$ and standard deviation $\sigma_M$ of the three 
measures $M$ were derived from the set of simulations.\\
As a pure frequentist approach,  we also considered the confidence or significance   
level of the  null hypothesis that the observation belongs to the Gaussian Monte Carlo 
ensemble  and  consider the fraction of simulations $p$, which 
have higher (lower) values of $M$ than the moments for the observation. 
In the Tables \ref{table1},  \ref{table2} and  \ref{table3}  the 
significances $S$ as well as the probabilities $p$ are listed.

%%%%%%%%%%%%%%%%%%%%%%%%%%%%%%%%
\begin{table}
\caption{$<\alpha>$ with Kp0 mask \label{table1}}

\begin{tabular}{lccc}
Scaling Range  &  Full Sky  &  Northern Sky & Southern Sky \\
 ($r$)  &  ($S$/\%)  &  ($S$/\%) & ($S$/\%) \\ \hline
0.025 & 2.2/99.5 & 2.1/99.4 & 2.3/99.8 \\
0.050 & 2.1/99.8 & 2.0/99.3 & 2.2/99.7 \\
0.075 & 2.0/99.5 & 1.9/99.3 & 2.0/99.4 \\
0.100 & 2.0/99.4 & 2.0/99.7 & 1.9/98.8 \\
0.125 & 2.0/99.5 & 2.1/99.8 & 1.9/98.6 \\
0.150 & 2.1/99.6 & 2.4/99.9 & 1.8/98.4 \\
0.175 & 2.2/99.7 & 2.5/$>$99.9 & 1.7/97.4 \\
0.200 & 2.2/99.7 & 2.7/$>$99.9 & 1.6/96.0 \\
0.225 & 2.2/99.7 & 2.7/$>$99.9 & 1.5/94.0 \\
0.250 & 2.1/99.4 & 2.8/$>$99.9 & 1.3/91.6 \\
\end{tabular}
\end{table}

%%%%%%%%%%%%%%%%%%%%%%%%%%%%%%%%

\begin{table} 
\caption{$\sigma_{\alpha}$ with Kp0 mask \label{table2}}

\begin{tabular}{lccc}
Scaling Range  &  Full Sky  &  Northern Sky & Southern Sky \\
 ($r$)  &  ($S$/\%)  &  ($S$/\%) & ($S$/\%) \\ \hline
0.025 & 2.0/98.9 & 1.9/98.7 & 1.9/98.9 \\
0.050 & 2.2/99.7 & 2.0/98.9 & 2.1/99.1 \\
0.075 & 0.4/64.8 & 0.2/60.3 & 0.4/63.8 \\
0.100 & 1.6/94.9 & 1.5/94.2 & 0.9/81.6 \\
0.125 & 2.2/99.0 & 2.6/99.8 & 1.0/84.6 \\
0.150 & 2.3/99.0 & 2.9/99.9 & 0.8/79.5 \\
0.175 & 2.1/98.4 & 2.9/99.9 & 0.6/71.6 \\
0.200 & 1.9/97.5 & 2.8/99.9 & 0.5/66.8 \\
0.225 & 1.7/95.8 & 2.6/99.7 & 0.4/64.3 \\
0.250 & 1.5/93.1 & 2.4/99.7 & 0.2/58.3 \\

\end{tabular}
\end{table}

%%%%%%%%%%%%%%%%%%%%%%%%%%%%%%%%

\begin{table} 
\caption{$\chi^2$ with Kp0 mask \label{table3}}

\begin{tabular}{lccc}
Scaling Range  &  Full Sky  &  Northern Sky & Southern Sky \\
 ($r$)  &  ($S$/\%)  &  ($S$/\%) & ($S$/\%) \\ \hline
0.025 & 2.0/96.7 & 1.8/95.9 & 2.1/96.9 \\
0.050 & 2.3/97.5 & 2.0/96.1 & 2.4/97.7 \\
0.075 & 0.9/88.4 & 0.7/85.9 & 1.0/88.4 \\
0.100 & 1.8/95.4 & 1.7/94.9 & 1.1/89.3 \\
0.125 & 2.7/97.8 & 3.6/98.9 & 1.0/89.2 \\
0.150 & 2.9/98.2 & 4.8/99.3 & 0.8/85.7 \\
0.175 & 2.7/97.9 & 5.2/99.4 & 0.5/80.3 \\
0.200 & 2.5/97.1 & 5.1/99.3 & 0.3/76.4 \\
0.225 & 2.1/96.6 & 4.8/99.3 & 0.1/70.1 \\
0.250 & 1.8/94.7 & 4.5/99.0 & 0.1/63.0 \\

\end{tabular}
\end{table}
%%%%%%%%%%%%%%%%%%%%%%%%%%%%%%%%

Analysing the significances and the confidence levels of the (combined) moments of 
the distribution of scaling indices  as a function of the scaling 
range the results read as follows: 
For the full sky we obtain for the mean value $<\alpha>$ significances ranging from  $2.0$ to $2.2$ 
for the different scales. The confidence levels for the detection of non-Gaussianties are, however,
very high and do not fall below $99 \% $ for any scale.
A closer look at the distribution of $<\alpha>$ for the simulations reveals 
that this distribution is not Gaussian but we find a number of outliers with very 
high values for $<\alpha>$, which leads to high $\sigma_{<\alpha>}$
thus to low significances. 
So although the significances for detecting 
non-Gaussianities are not very high, the confidence levels, for which no implicit assumptions
about the distribution function of the test statistics is made,  strongly indicate
the presence of non-Gaussian features in the observed WMAP-data.
Even higher values for both the significances and the confidence levels are found, if one only considers
the northern hemisphere. In this case the  sigificances range from  $1.9$ for smaller scales  to $2.8$ for
the largest scale. For scales larger than $r=0.15$ none of the simulations was found to have a higher values for $<\alpha>$
than the observation, which represents a quite unambigous detection 
of non-Gaussianties in the northern hemisphere.   
For the southern hemisphere, however, both the significances and the confidence levels for the 
smaller radii are slighly higher than for the northern sky but continously decrease 
for higher radii $r$. 
For the standard deviation we find slighly different results. For the smallest scales ($r=0.025$) 
$\sigma_{\alpha}$ is significantly larger for WMAP than for the simulations. In a transition regime  
$r \approx 0.075$ the standard deviation is practically the same for the observation and the Monte Carlo sample.
For larger scales we observe higher standard deviations for the simulations.  
This effect is much more pronounced in the northern hemisphere giving rise to significances 
up to $2.9$ and very high confidence levels for intermediate scales. For the largest scales the differences for $\sigma$
between simulations and oberservation diminishes. 
For the southern hemisphere the width of the distributions  become more and more similar so that 
no signatures for deviations from Gaussianity are identified using $\sigma_{\alpha}$. 
The behaviour of the $\chi^2$- statistics as a function of the scale parameters $r$  can -- as expected -- be regarded as
a superposition of the two underlying statistics $<\alpha>$ and $\sigma_{\alpha}$.
Only the significances are highly increased leading to a $5 \sigma$ detection of non-Gaussianity 
at scales of $r=0.175$ in the northern sky.
Performing the same analyses for the extended mask (Figure  \ref{fig12}) yields essentially the same results. 
Only marginal variations for the  significances and confidence levels (which are not explicitly shown in this paper) 
are found in this case.
Thus, these results  are quite stable with respect to differences for the chosen mask.\\ 
Some readers might argue that the selection of the moments and especially of the scales, where the significances are largest,
represents an {\it a posteriori} choice taken after looking at the data.
However, the choice of the moments was motivated by the results obtained with simulations (see Section 3) and
the investigation of the data on different length scales can be regarded as a fairly conventional and unbiased  approach. 
If a random field is Gaussian, it must be Gaussian on all scales. If deviations are detected at some scales, one can already 
infer non-Gaussianity. 
Nevertheless, we also calculated diagonal  $\chi^2$-statistics, where we considered only one 
(mean or standard deviation) or both
measures, and {\it summed over all length scales} ,
\begin{equation}
\chi^2_{<\alpha>}= \sum_{i=1}^{N_r}   \left[ \frac{M_1(r_i) - <M_1(r_i)>}{\sigma_{M_1(r_i)}} \right]^2
\end{equation}

\begin{equation}
\chi^2_{\sigma_{\alpha}}= \sum_{i=1}^{N_r}   \left[ \frac{M_2(r_i) - <M_2(r_i)>}{\sigma_{M_1(r_i)}} \right]^2
\end{equation}

\begin{equation}
\chi^2_{<\alpha>,\sigma_{\alpha}}=  \sum_{i=1}^{N_r} \sum_{j=1}^2  \left[ \frac{M_j(r_i) - <M_j(r_i)>}{\sigma_{M_j(r_i)}} \right]^2
\end{equation}

with $M_1(r_i)=<\alpha(r_i)>$, $M_2(r_i)=\sigma_{\alpha}(r_i)$ and $N_r$ being the number of considered length 
scales (here: $N_r=10$). 
The results are summarised in Table \ref{table4}.
%%%%%%%%%%%%%%%%%%%%%%%%%%%%%%%%%%%%%%

\begin{table} 
\caption{$\chi^2$,  all scales \label{table4}}

\begin{tabular}{lccc}
 $\chi^2$ &  Full Sky  &  Northern Sky & Southern Sky \\
        &  ($S$/\%)  &  ($S$/\%) & ($S$/\%) \\ \hline
$\chi^2_{<\alpha>}$ & 2.1/96.9 & 2.7/97.7 & 1.5/94.2 \\
$\chi^2_{\sigma_{\alpha}}$ & 2.4/96.5 & 4.4/99.5 & 0.2/70.0 \\
$\chi^2_{<\alpha>, \sigma_{\alpha}}$ & 2.4/97.3 & 3.7/98.9 & 1.1/91.6 \\
\end{tabular}
\end{table}

%%%%%%%%%%%%%%%%%%%%%%%%%%%%%%%%%%%%%%

\begin{figure}
\centering
 \includegraphics[scale =0.5,angle=0]{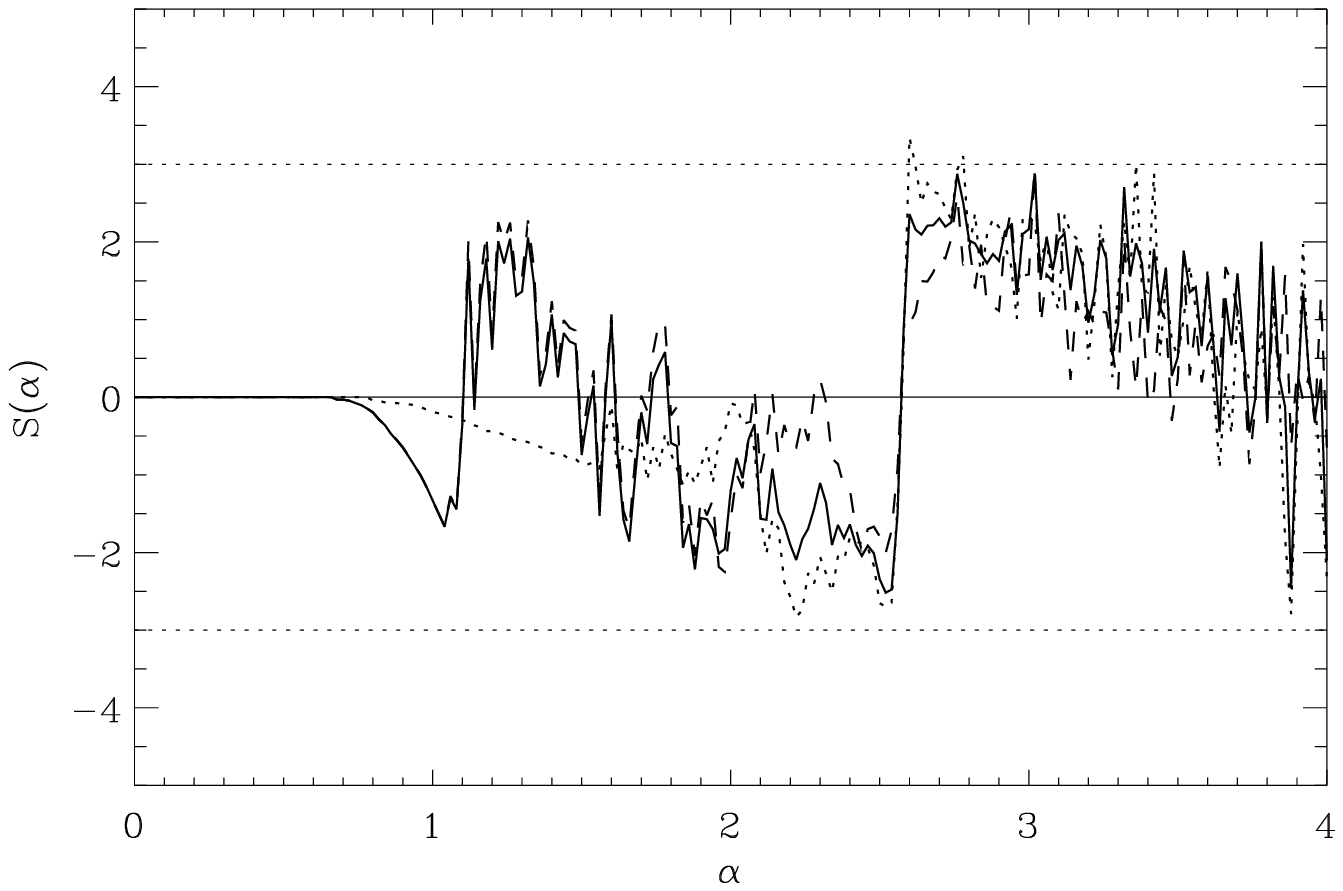}
  \includegraphics[scale =0.5,angle=0]{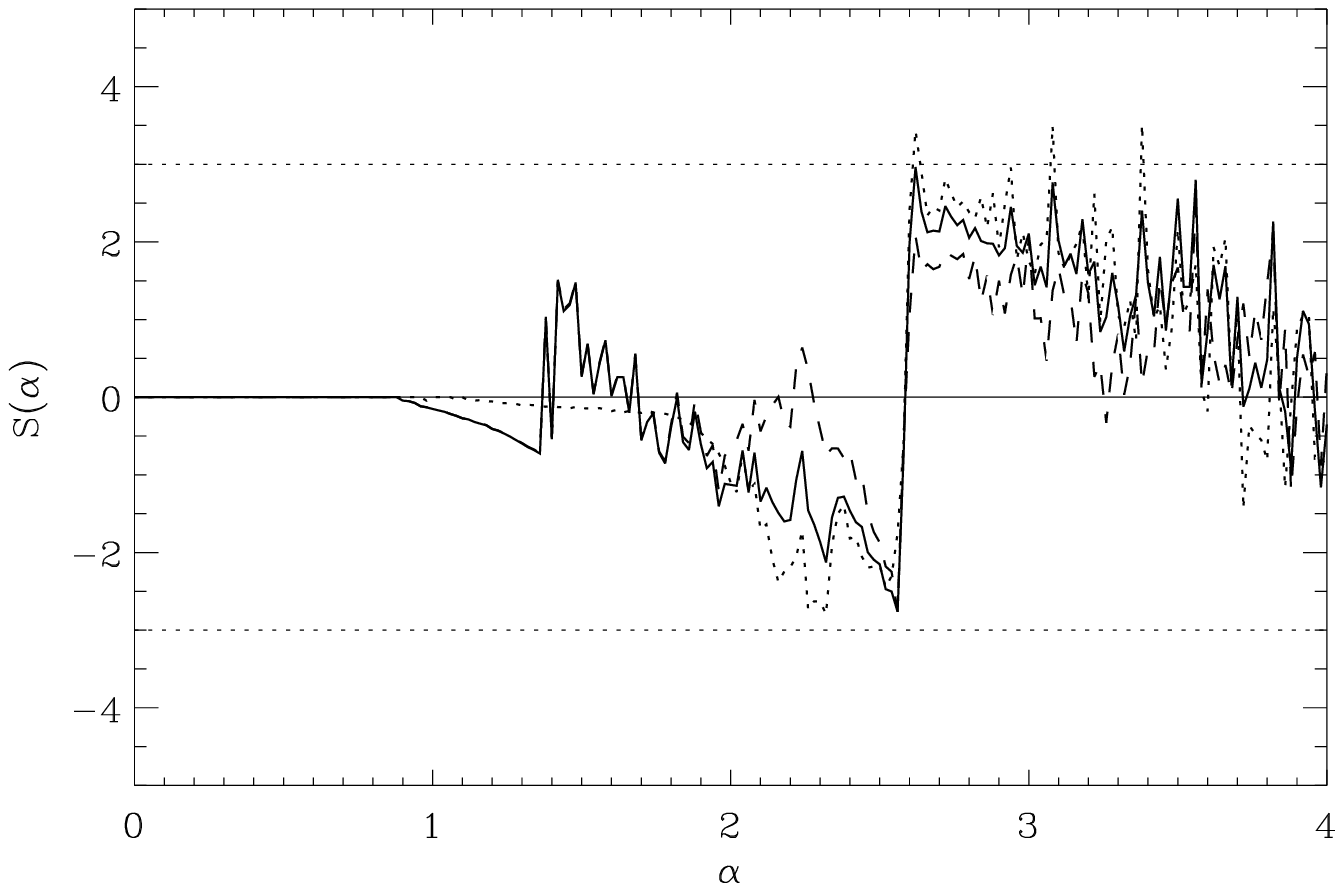}
\caption{Bin-wise significances (with sign) as derived from the pribability distribution
              $P(\alpha)$ for the WMAP data and Kp0-mask for $r=0.175$ (above) and $r=0.225$ (below). \label{fig13}}
\end{figure}

\begin{figure}
\centering
 \includegraphics[scale =0.5,angle=0]{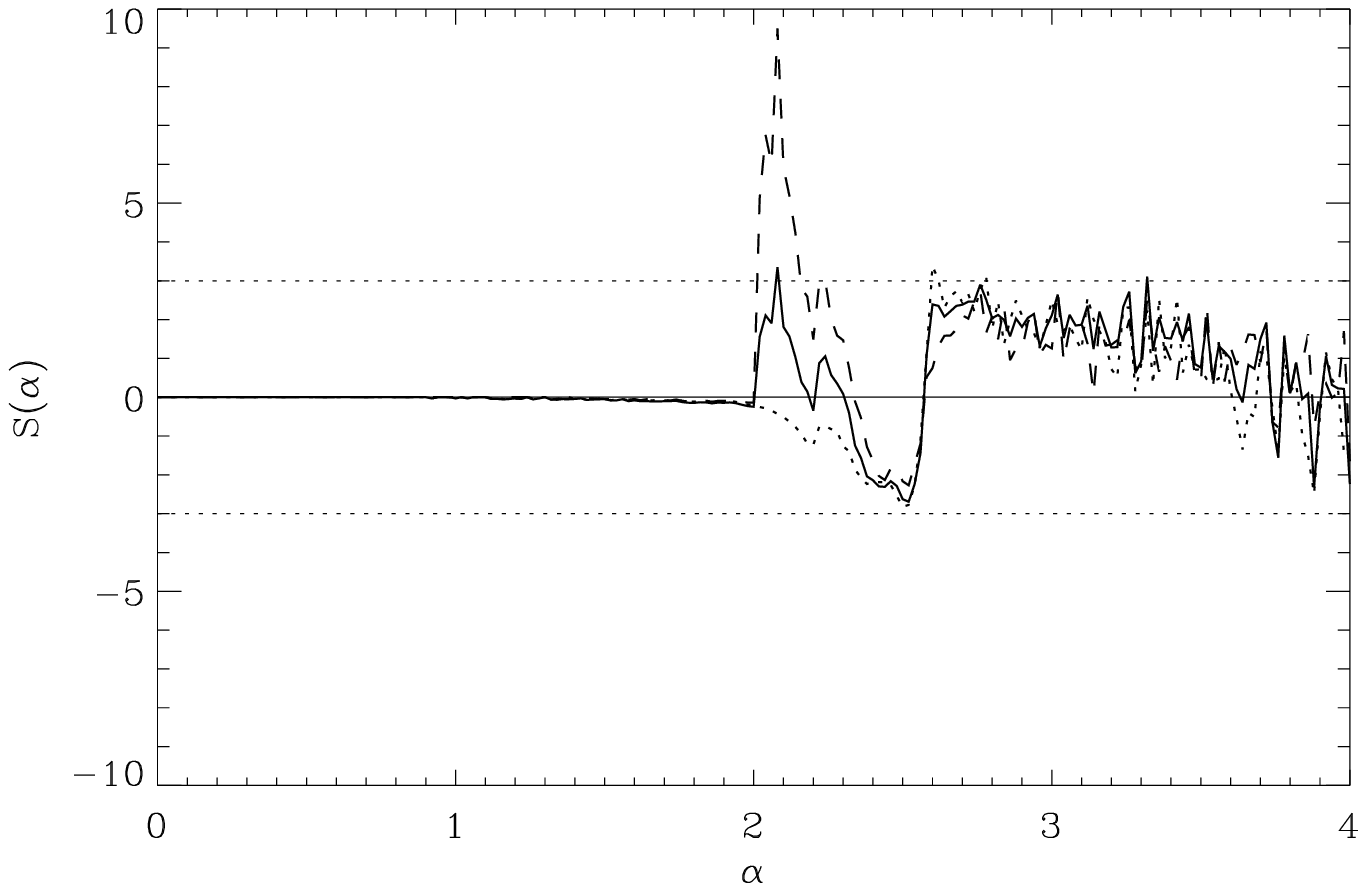}
  \includegraphics[scale =0.5,angle=0]{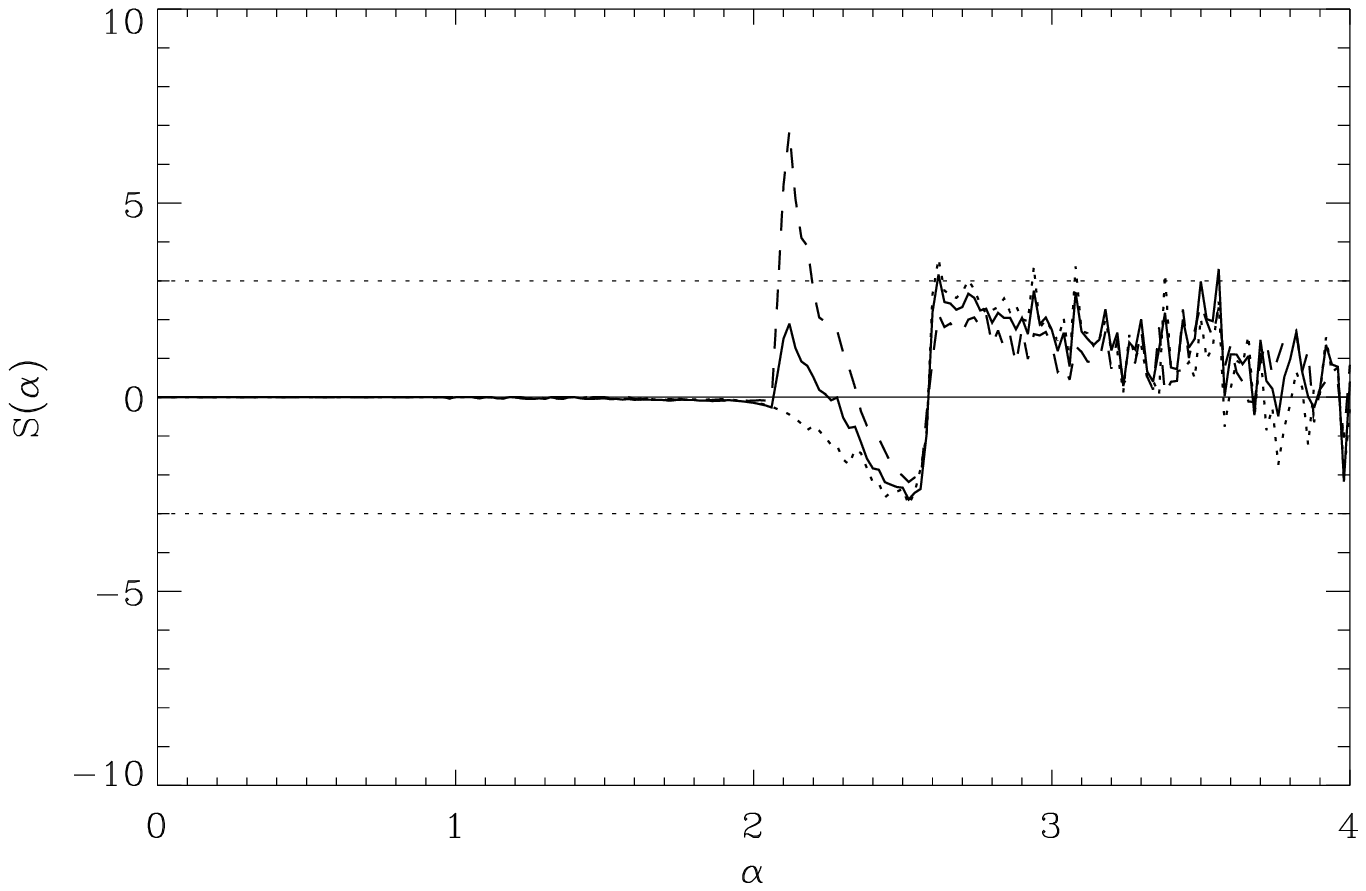}
\caption{Same as fig. \ref{fig13} but for the extended mask. \label{fig14}}
\end{figure}

Also for this obviously a priori test statistics, where some unimportant scales contribute to the final 
value of $\chi^2$  and may dilute the result, we find significant signatures for non-Gaussianities in the northern sky.\\
Beside the previous analyses based on global measures derived from the $P(\alpha)$- distribution, 
we also investigate the spectrum of scaling indices on a differential level.
Therefore we consider the bin-wise significances  $S(\alpha_i)$,
\begin{equation}
S(\alpha_i) = \ \frac{P(\alpha_i) - <P(\alpha_i)>}{\sigma_{P(\alpha_i)}} 
\end{equation}
for the probability densities $P(\alpha)$ of selected scales (see Figures  \ref{fig13} and  \ref{fig14}). 

%%%%%%%%%%%%%%%%%%%%%%%%%%%%%%%%
%%%%%%%%%%%%%%%%%%%%%%%%%%%%%%%%
\begin{figure}
\centering
 \includegraphics[scale =.5,angle=0]{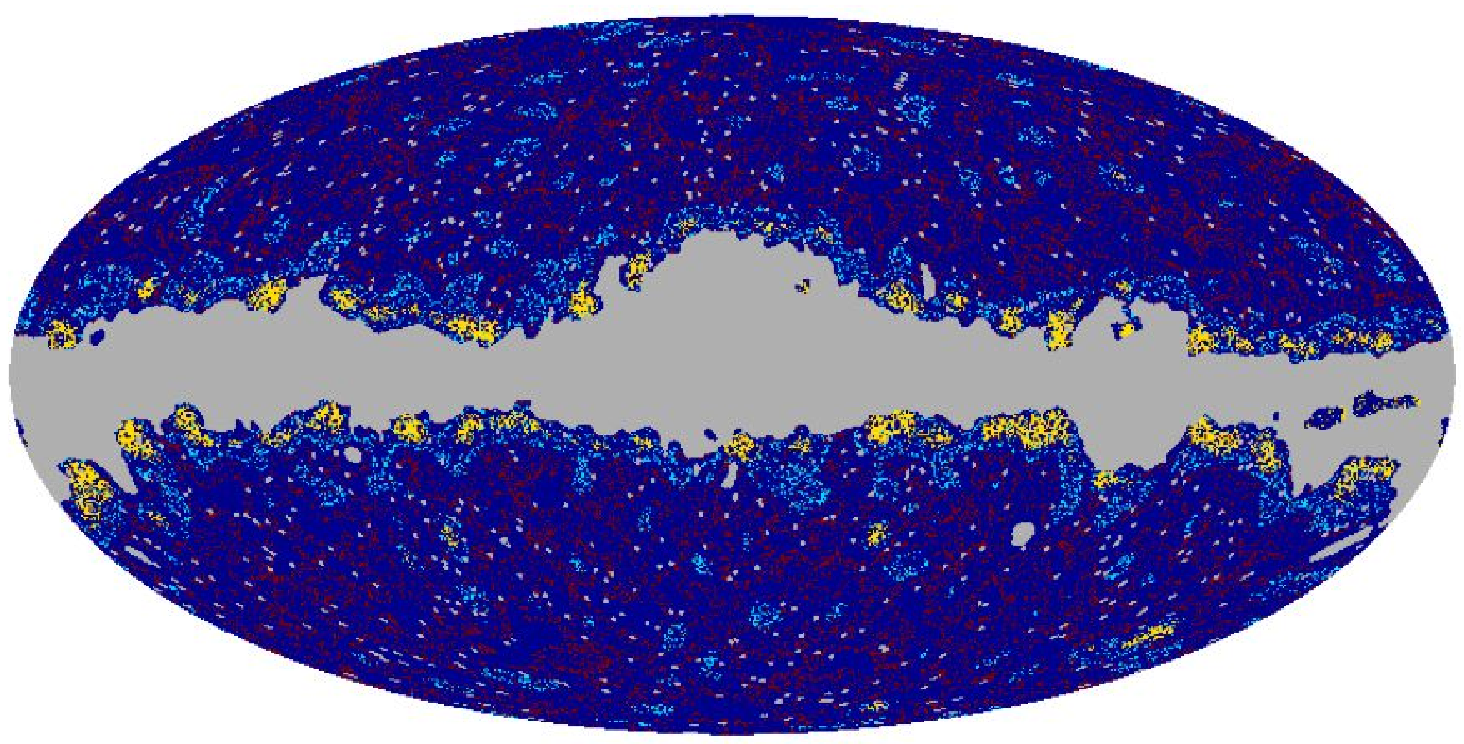}
  \includegraphics[scale =.5,angle=0]{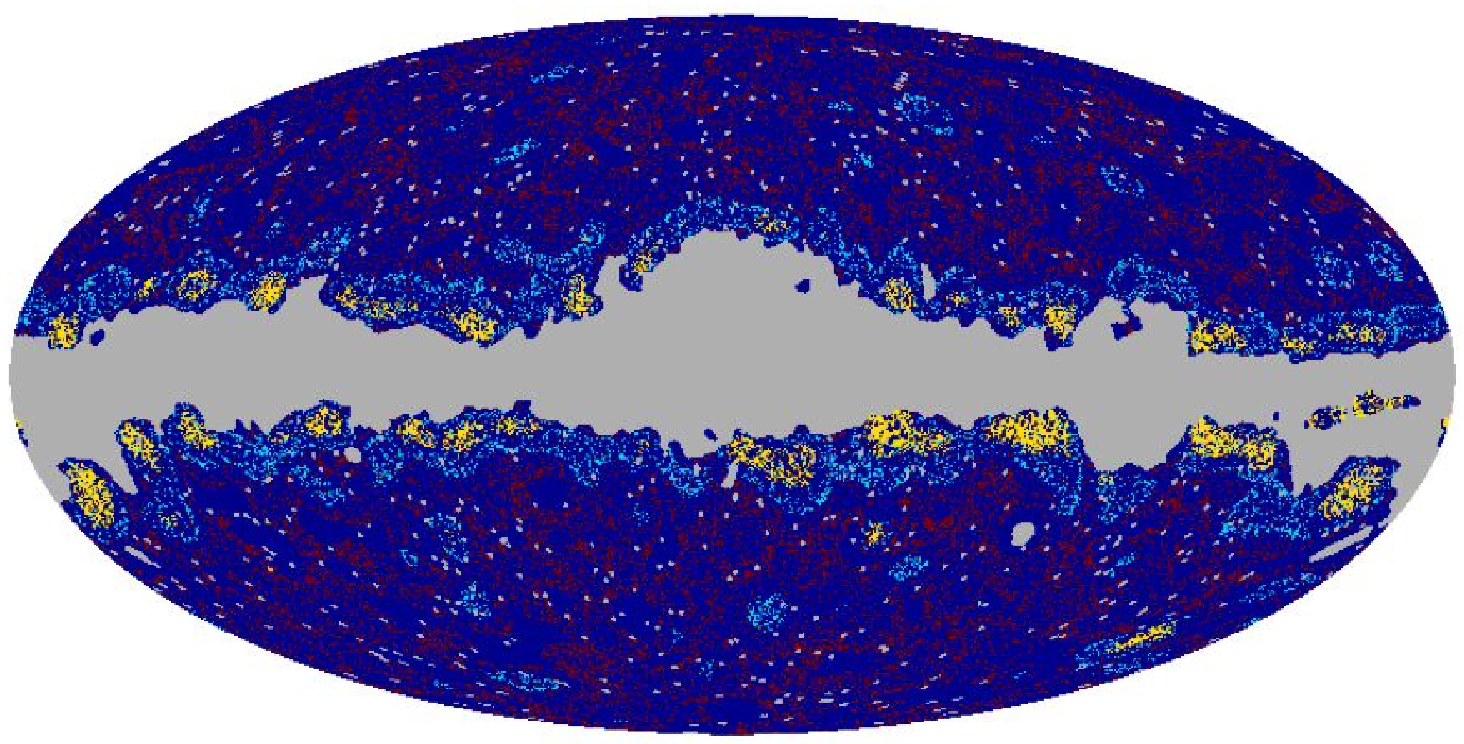}
\caption{Marked pixels with $\alpha \in [2.0,2.3]$ (yellow), $\alpha \in [2.425,2.475]$ (blue) 
              and $\alpha \in [2.60,2.65]$ (red) for $r=0.175$ (above) and $r=0.225$ (below). \label{fig15}}
\end{figure}

%%%%%%%%%%%%%%%%%%%%%%%%%%%%%%%%
Note that we omitted in this case the absolute value in the definition of the significance 
in order to show in which direction with respect to the simulations the $P(\alpha)$ of the WMAP data
deviates. 
Although the spectra for the Kp0-mask (Figure  \ref{fig13}) are diluted by edge effects 
for smaller $\alpha$, one can clearly 
detect the sharp transition from systematically small to high values in the significances, 
which is due to the shift of the whole spectrum 
towards higher $\alpha$-values  for the WMAP-data. 
This shift is so pronounced that for some bins of the spectrum even the $3\sigma$- level 
for the deviation is nearly reached. 
If we consider the same spectra for the extended mask, where almost all disturbing edge effects are 
removed,  we see (Figure  \ref{fig14}) - beside the transition from negative to positive values due to the global shift - 
a new, highly significant feature at $\alpha \approx 2.1$ in the southern sky emerging, when the 
scaling range is increased. There is an excess of pixels with these low $\alpha-$values in the WMAP-data 
as compared to the simulated maps.
To elucidate the origin of these pixels we color-coded the pixels according their value of scaling indices (Figure  \ref{fig15}). 
%%%%%%%%%%%%%%%%%%%%%%%%%%%%%%%%
%%%%%%%%%%%%%%%%%%%%%%%%%%%%%%%%

%%%%%%%%%%%%%%%%%%%%%%%%%%%%%%%%
\begin{figure}
\centering

 \includegraphics[scale =0.5,angle=0]{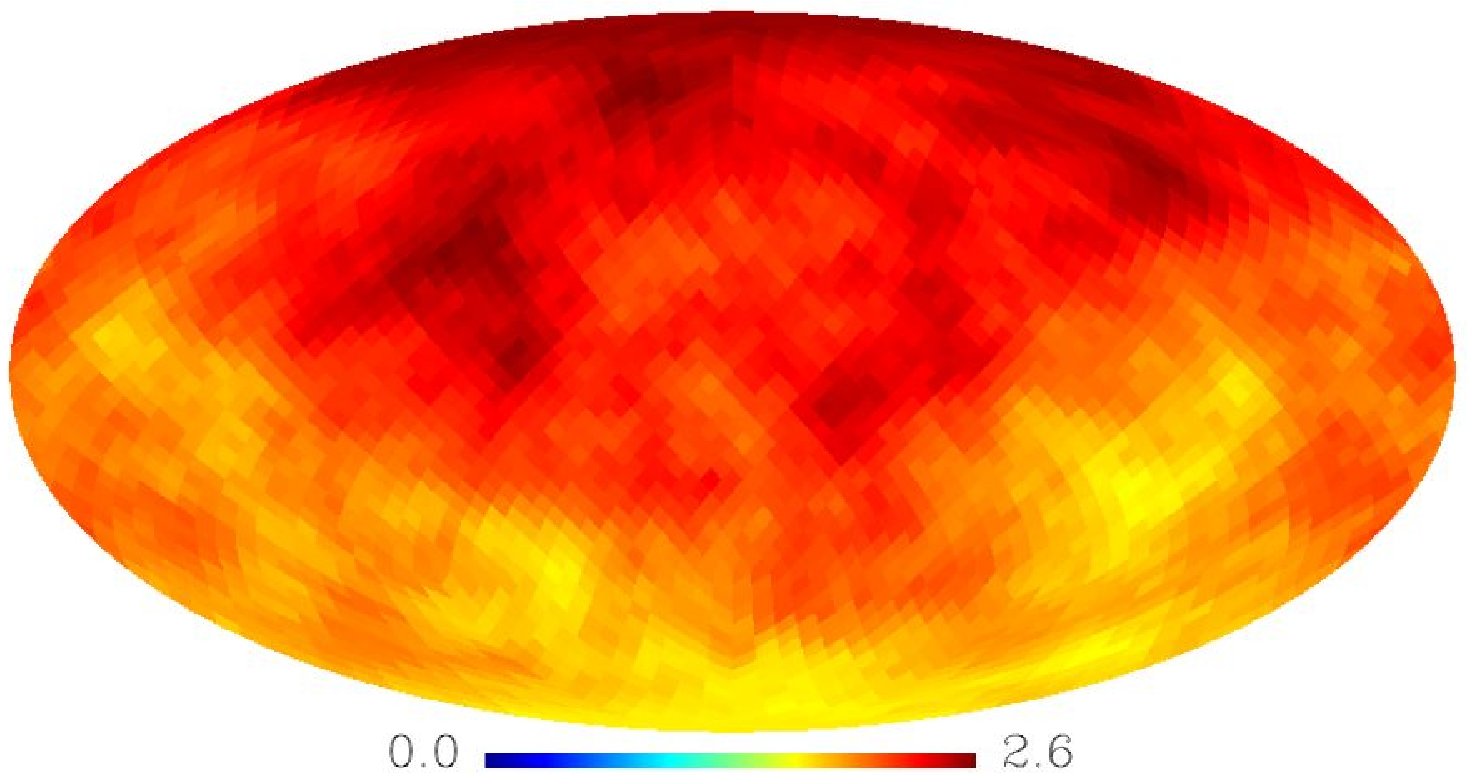}
 \includegraphics[scale =0.5,angle=0]{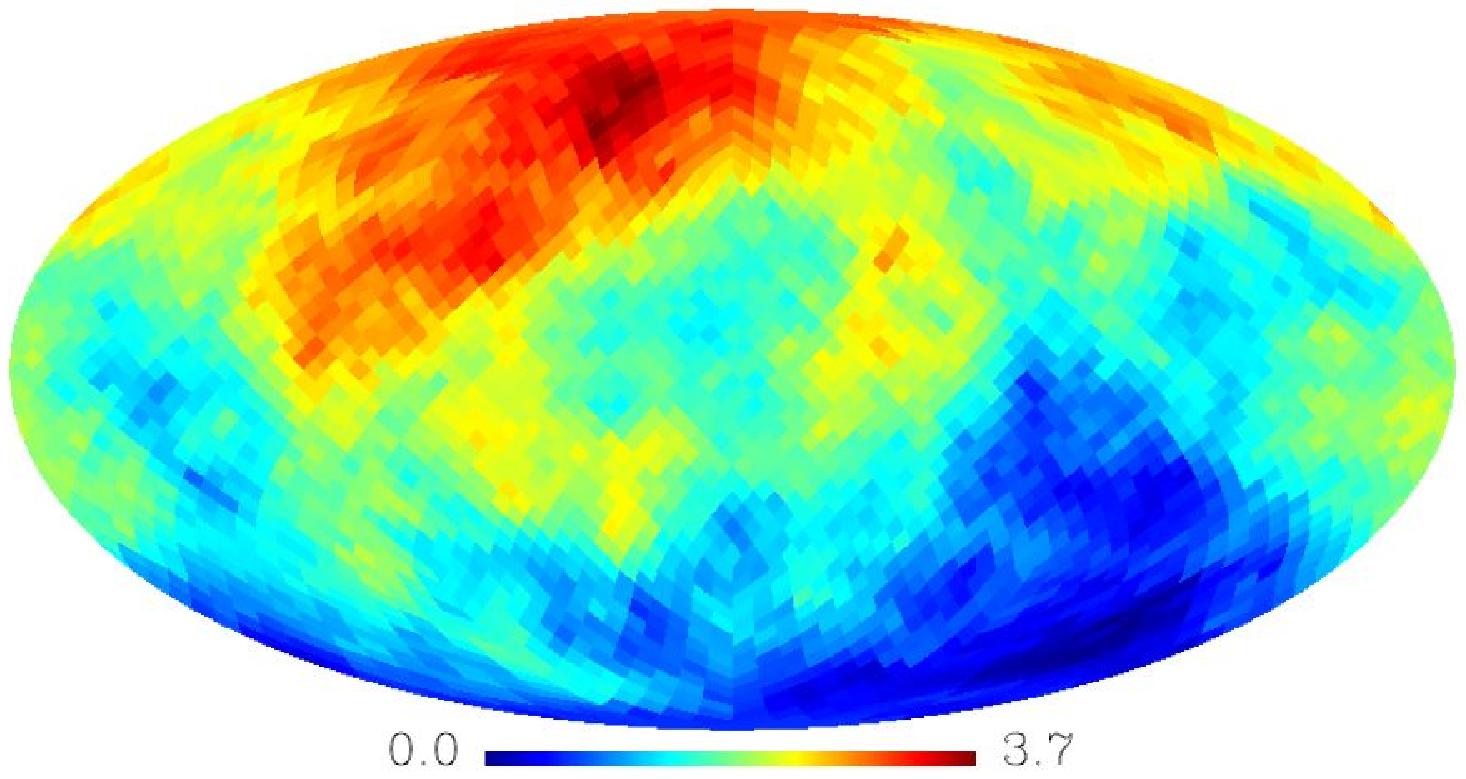}
  \includegraphics[scale =0.5,angle=0]{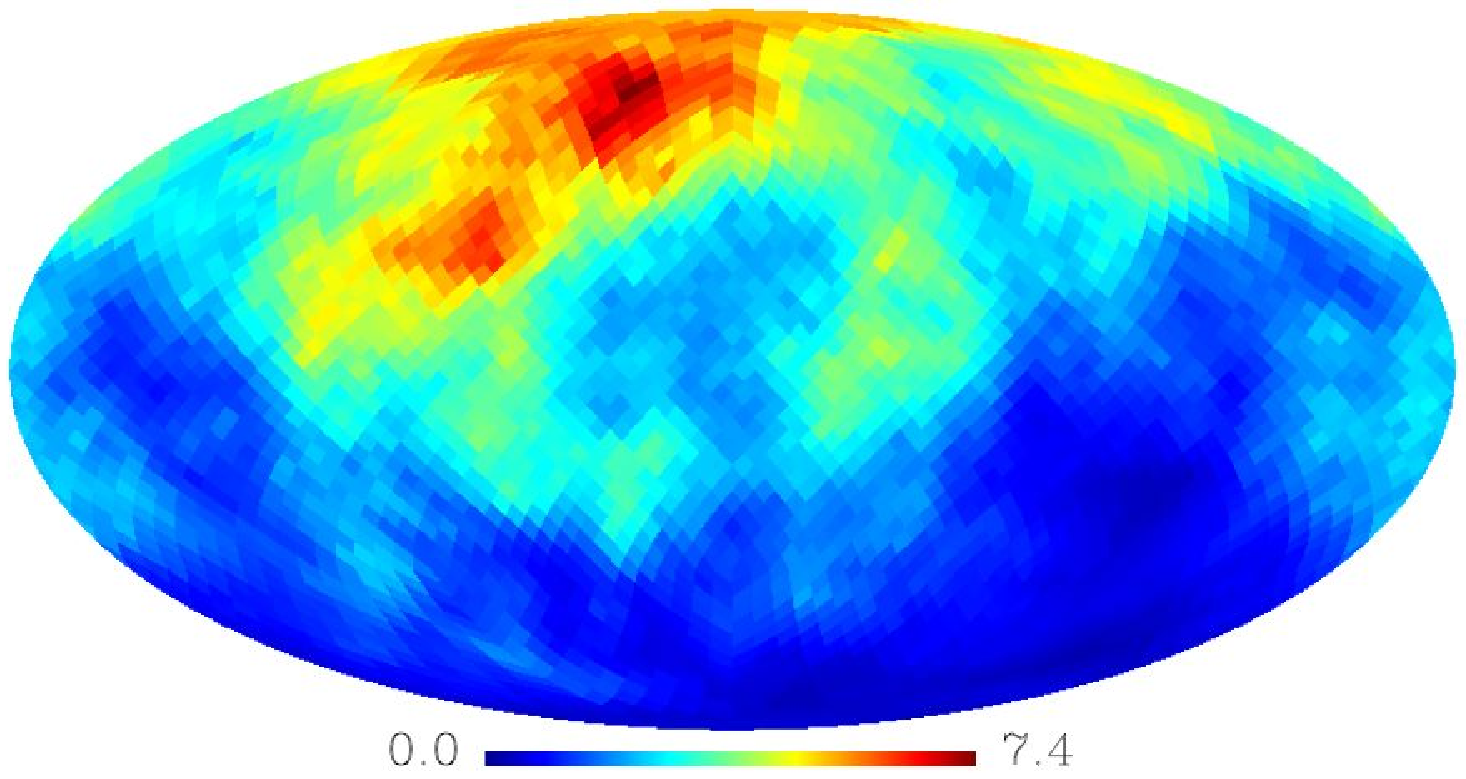}
\caption{Significanes of the (rotated) upper hemisphere for the SIM based statistics for $r=0.175$. 
              From top to bottom: mean $<\alpha>$, standard deviation $\sigma_{\alpha}$ and $\chi^2$-combination of
              mean and standard deviation. The highest significances were obtained for 
              $(\theta,\phi)= (27^{\circ},35^{\circ})$ for $<\alpha>$, $(\theta,\phi)= (39^{\circ},45^{\circ})$ 
              for $\sigma_{\alpha}$ and  $(\theta,\phi)= (30^{\circ},41^{\circ})$ for $\chi^2$.
              The maximal asymmetry between the rotated northern and southern hemisphere were found 
              for the rotation angles $(\theta,\phi)= (27^{\circ},35^{\circ})$ for $<\alpha>$, $(\theta,\phi)= (30^{\circ},41^{\circ})$ 
              for $\sigma_{\alpha}$ and  $(\theta,\phi)= (30^{\circ},41^{\circ})$ for $\chi^2$.   \label{fig16}}
\end{figure}

%%%%%%%%%%%%%%%%%%%%%%%%%%%%%%%%
\begin{figure}
\centering
 \includegraphics[scale =0.5,angle=0]{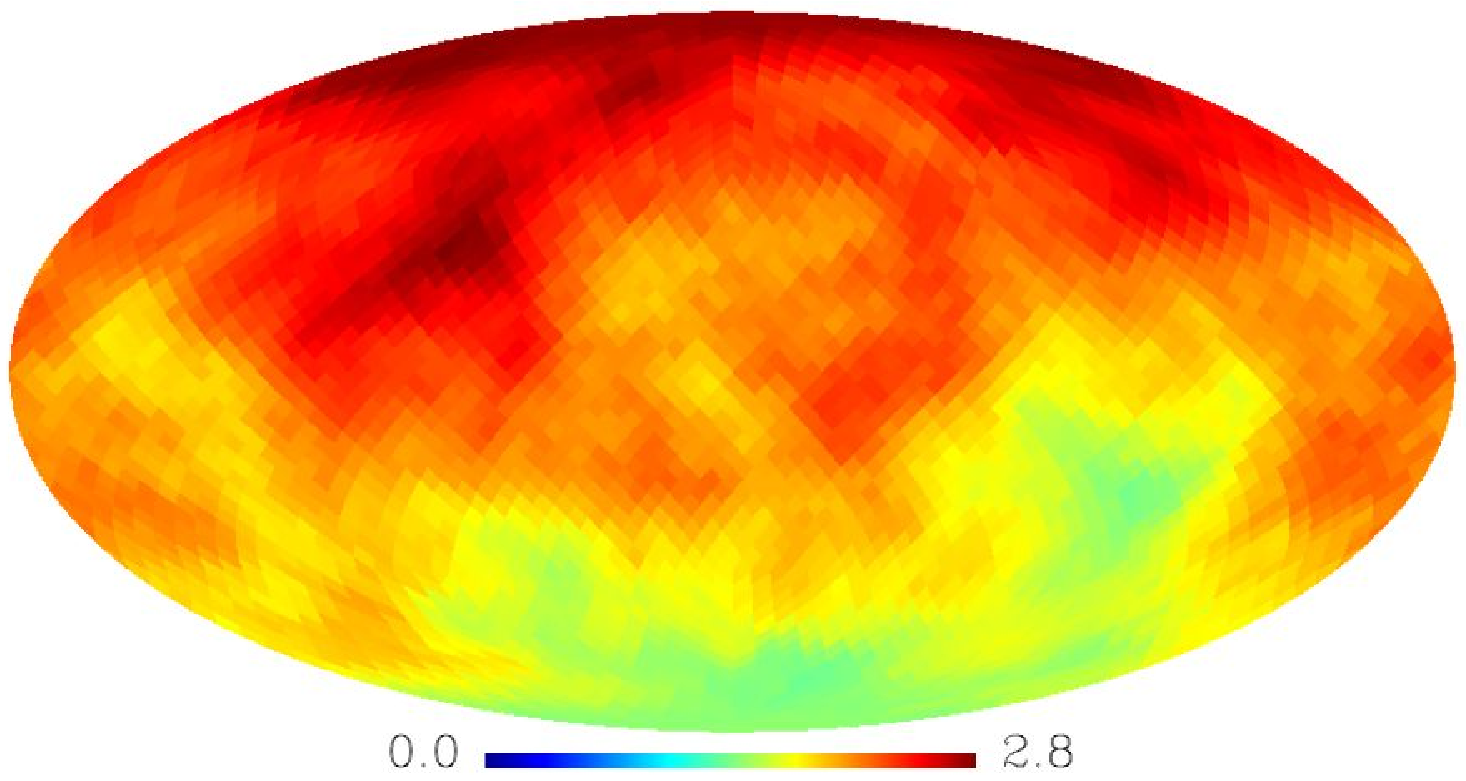}
  \includegraphics[scale =0.5,angle=0]{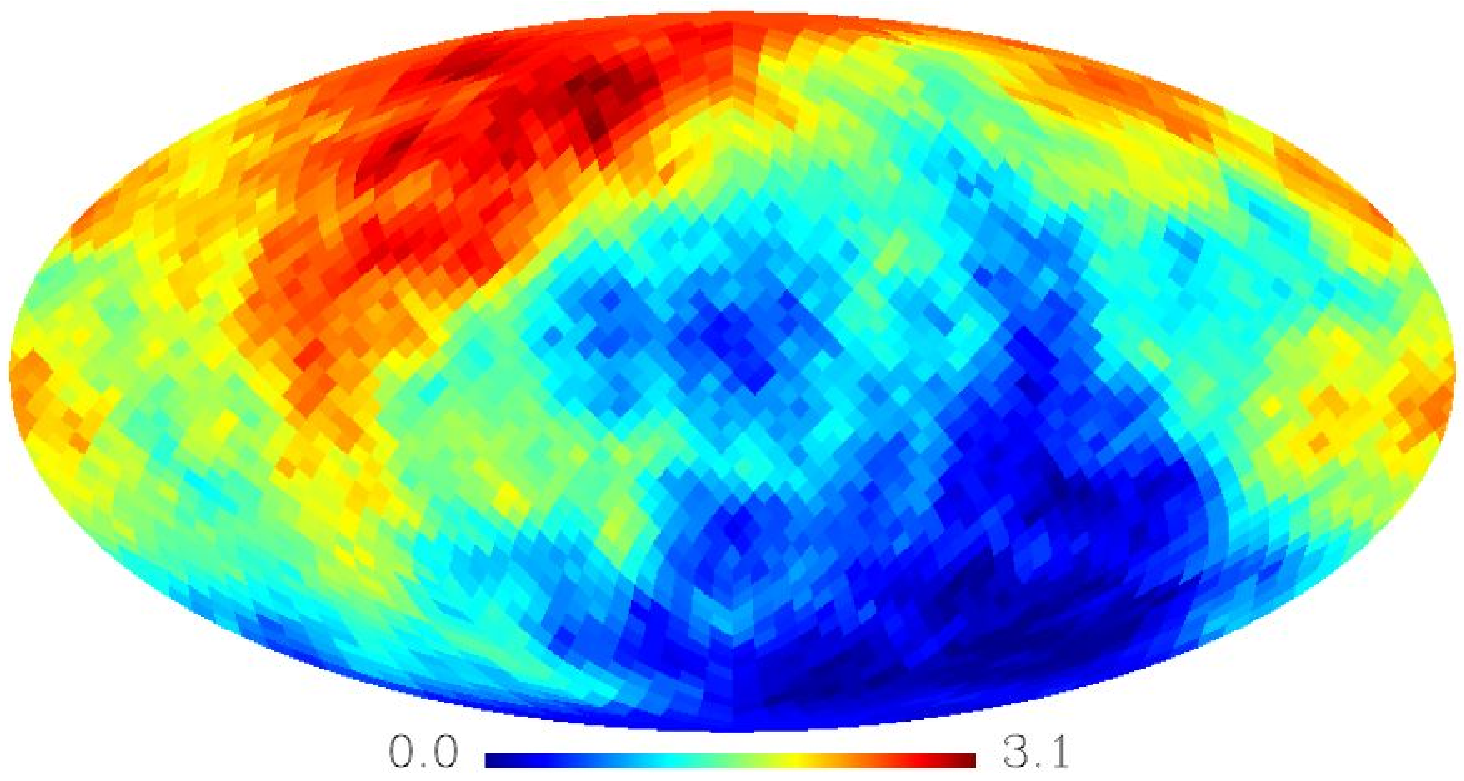}
  \includegraphics[scale =0.5,angle=0]{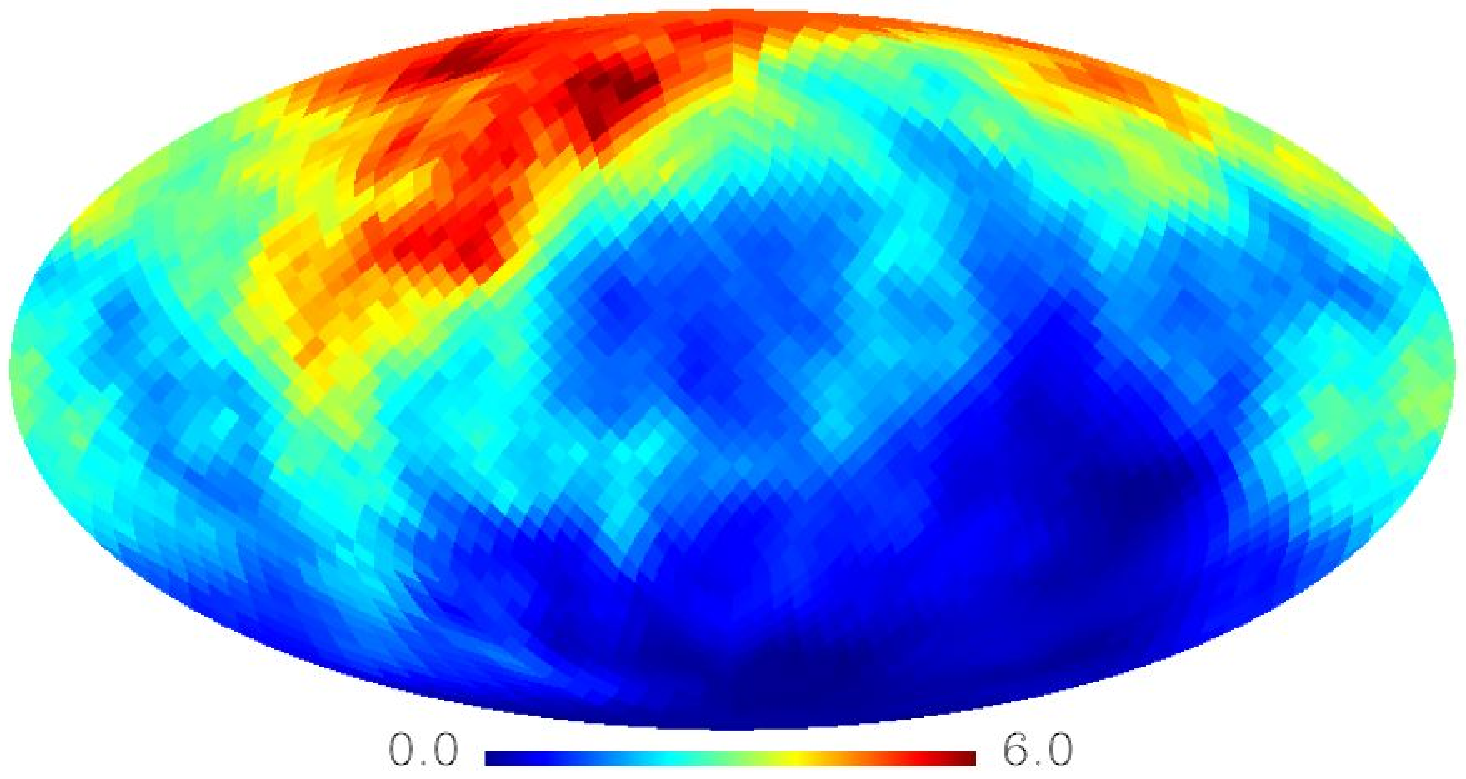}
\caption{Same as Fig. \ref{fig16} but for $r=0.225$. For this scaling range, the highest significances 
              were obtained for 
              $(\theta,\phi)= (21^{\circ},148^{\circ})$ for $<\alpha>$), $(\theta,\phi)= (27^{\circ},65^{\circ})$ 
              for $\sigma_{\alpha}$ and  $(\theta,\phi)= (30^{\circ},41^{\circ})$ for $\chi^2$.
              The maximal asymmetries between the rotated northern and southern hemisphere were found 
              for the rotation angles $(\theta,\phi)= (21^{\circ},148^{\circ})$ for $<\alpha>$, $(\theta,\phi)= (27^{\circ},55^{\circ})$ 
              for $\sigma_{\alpha}$ and  $(\theta,\phi)= (30^{\circ},41^{\circ})$ for $\chi^2$.   \label{fig17}}
\end{figure}
%%%%%%%%%%%%%%%%%%%%%%%%%%%%%%%%%%
\clearpage

It becomes immediately obvious that the pixels with the small $\alpha$-values identified in the southern sky 
form  two to three clusters (depending on the scale $r$), where the largest one 
corresponds to the cold spot --  the well-known 
anomaly first detected by \citet{vielva04a} by means of a wavelets analysis.  
The pixels  belonging to the other two $\alpha$-intervals, for which the largest deviations in the bulge of the 
distribution are found, cannot be associated with special localised features. These pixels are rather 
distributed all over the sky, which indicates that the shift of the spectrum of scaling indices  
represents a global intrinsic effect.\\ 
The distinction between northern and southern hemisphere, which we made during the previous 
analysis, is natural being triggered by excluding the foreground-contamined 
area of the galactic plane.
On the other hand, this choice is arbitrary, because no symmetry axis is preferable for a 
presumably isotropic CMB. 
To test for asymmetries in WMAP-data we consider rotated hemispheres, calculate the 
global measures $<\alpha>, \sigma_{\alpha}$ and $\chi^2$ for  $3072$ rotation angles and
compare the results for the (rotated) northern and southern hemispheres.
In Figures  \ref{fig16} and  \ref{fig17}  we show the 
significances $S(<\alpha>)$, $S(\sigma_{\alpha})$ and  $S(\chi^2)$  of the 
northern hemisphere as determined in a reference frame where the north pole 
pierces the center of the color-coded pixel.   
All three measures yield systematically higher significances for rotations pointing to northern directions 
relative to the galactic coordinate system. 
The rotation angles, for which the highest significances of the 
northern hemisphere  are obtained, 
are  listed in the captions of the respective figures. 
They are very similar but do not always coincide with the rotations, 
for which the maximal asymmetries, as measured by the 
difference of the significances of the rotated northern and southern 
hemisphere are found.
It is worth noticing that the  direction of the most pronounced non-Gaussianities and asymmetries differs 
from the dipole direction $(\theta, \phi)=(42^{\circ}, 264^{\circ} )$ and the
so-called axis of evil  $(\theta, \phi)=(30^{\circ},260^{\circ})$, which is very close to the dipole direction.

\section{Summary}
 
We performed a scaling index analysis of the WMAP three-year data. Specifically, 
we analysed the foreground-cleaned co-added maps of the V-and W-band. 
We found highly significant signatures of both non-Gaussianities and asymmetries in the WMAP 
three-year data. Our main findings can be summarized and interpreted as follows:\\ 
In the northern hemisphere the spectrum of scaling indices is systematically broader and 
shifted towards higher values yielding highly significant deviations of the mean and 
standard deviation of the distribution. 
This effect can naturally be interpreted as too few structure and structural variations 
in the CMB-fluctuations as measured by WMAP compared to the predicted ones 
within the concordance model.
The highest global signatures for non-Gaussianities and asymmetries 
between the northern and southern hemisphere were found for rotated coordinate systems,
where the significances for the detection of signatures for non-Gaussianities
range from $2.6 \sigma$ up to $7.4 \sigma$.
These findings are quite consistent with previous results for the first year 
data \citep{eriksen04a, eriksen04b, park04a}, 
where very similar features were identified -- even though with a smaller 
significance level -- using 
the Minkowski functionals, the power spectrum and $N$-point 
correlation functions.\\
In the southern hemisphere the global properties of the $P(\alpha)$ distribution 
for the WMAP-data are more consistent with the simulations than in the northern sky.
We find, however, highly significant localisable features of non-Gaussianity, for which 
the largest one can be associated with a cold spot. An anomalous  area  detected 
by \citet{vielva04a} in the WMAP first year data and confirmed in the WMAP 
three year data \citep{cruz06a}.

\section{Conclusions}
 
In conclusion, we demonstrated for the first time the feasibility to adapt and apply 
the scaling index method as an estimator of local scaling properties of a point set
to spherical symmetric data.\\
The results obtained with the scaling indices give further strong evidence 
that also the coadded WMAP three year data do indeed contain 
unusual features, which are not in agreement with the 
hypotheses of Gaussianity and isotropy
predicted in the standard inflationary scenario. \\
A quite remarkable result of our study is the  fact that 
two previously known anomalies, namely the lack 
of structure in the northern sky and the cold spot in the southern hemisphere
could be reidentified using a completely different test statistic.
This increases the evidence that these anomalies are of true physical origin and not
related to one single test statistic.\\
It is also worth noticing that the scaling indices could detect both of these 
anomalies simultaneously, whereas the wavelets were only sensitive to localised 
spot-like structures and the $N$-point correlation functions and Minkowski-functionals
only detected the non-Gaussianities and asymmetries on large scales.\\   
The main task for future studies is to elucidate the possible sources of these anomalies, 
whether they are due to systematics or foreground effects or truely represent intrinsic
CMB-fluctuations due to some exotic physics.  
Future more detailed studies will investigate the possible origins of
the non-Gaussian signatures by separately analysing the V and W bands and  by
comparing the three year WMAP-data with the first, second and third year data.
All these tests may help to find hints about the origin of the non-Gaussian signatures.\\
In either case we could demonstrate that the scaling index method provides very flexible 
and highly sensitive statistics for e.g. the identification of asymmetries and non-Gaussian signatures 
in the WMAP-data thus representing an important novel statistical tool for CMB analyses.

\section*{Acknowledgments}

Many of the results in this paper have been derived using the HEALPix \citep{gorski05a} software 
and analysis package. We acknowledge use of
the Legacy Archive for Microwave Background Data Analysis (LAMBDA). 
Support for LAMBDA is provided by the NASA Office of Space Science.\\
This work is dedicated to Peter Schuecker, who suddenly deceased -- much too soon  -- during
the preparation of this manuscript. We (will) miss him a lot.

\label{lastpage}

\end{document}